\documentclass[11pt,A4Paper]{article}
\usepackage{fullpage}
\usepackage{hyperref}
\usepackage{graphicx}
\usepackage[labelsep=period]{caption}
\usepackage{authblk}
\usepackage{rotating}
\usepackage{xcolor}
\usepackage{ulem}

\usepackage[utf8]{inputenc}
\usepackage{amsmath}
\usepackage{amsfonts}
\usepackage{gensymb}

\def\Xint#1{\mathchoice
   {\XXint\displaystyle\textstyle{#1}}%
   {\XXint\textstyle\scriptstyle{#1}}%
   {\XXint\scriptstyle\scriptscriptstyle{#1}}%
   {\XXint\scriptscriptstyle\scriptscriptstyle{#1}}%
   \!\int}
\def\XXint#1#2#3{{\setbox0=\hbox{$#1{#2#3}{\int}$}
     \vcenter{\hbox{$#2#3$}}\kern-.5\wd0}}

\def\dashint{\Xint-}

\newcommand{\no}{\noindent}

\newcommand{\beq}{\begin{equation}}
\newcommand{\eeq}{\end{equation}}

\title{Stokes Waves in Water of Finite Depth}
\author[1]{\Large Eleanor Byrnes\thanks{Corresponding author}}
\author[2]{\Large Bernard Deconinck}
\author[3]{\Large Anastassiya Semenova}
\affil[$$]{\textit{\small Department of Applied Mathematics, University of Washington, Seattle, WA 98195-3925}}
\affil[$$]{\textit{\small $^1$elbyrnes@uw.edu, $^2$deconinc@uw.edu, $^3$asemenov@uw.edu}}
\date{\today}
\begin{document}

\maketitle

\begin{abstract}
Periodic water waves of permanent form traveling at constant speed, the so-called Stokes waves, are studied in water of fixed finite depth using methods previously used in water of infinite depth. We apply our methods to waves of varying steepness over a range of fixed depths in order to determine how a number of physical quantities related to the waves change as the steepness of the waves increases. Finally, we examine the Taylor sign condition for these waves, as well as the complex singularities outside of their domain of definition when the waves are considered as a function of a conformal variable. 

%We modify the method of  \cite{semenova2024stokes} to compute periodic water waves of permanent form, traveling at constant speed the so-called Stokes Waves, in water of fixed finite depth. We apply this method to waves of varying steepness over a range of fixed depths in order to determine how a number of quantities related to the waves change, as the steepness of the waves increases.
\end{abstract}

\section{Introduction}

The two-dimensional Euler water wave problem describes the evolution of a body of an incompressible, inviscid, and irrotational fluid with a one-dimensional free surface under the effect of gravity. Assuming surface tension to be negligible, the Euler water wave problem is water of depth $d$ is

\begin{eqnarray}
\nabla^2\phi=0,&& ~~~~-d<y<\eta(x,t),\label{eq:Laplace}\\
\phi_y=0, && ~~~~ y=-d,\label{eq:noPenetration}\\
\eta_t+\phi_x\eta_x=\phi_y, && ~~~~y=\eta(x,t),\label{eq:KinematicEuler}\\
\phi_t+g\eta+\frac{1}{2}\left(\phi_x^2+\phi_y^2\right)=
0, && ~~~~y=\eta(x,t),\label{eq:dynamicEuler}
\end{eqnarray}

\begin{figure}
    \centering
    \includegraphics[width=\linewidth]{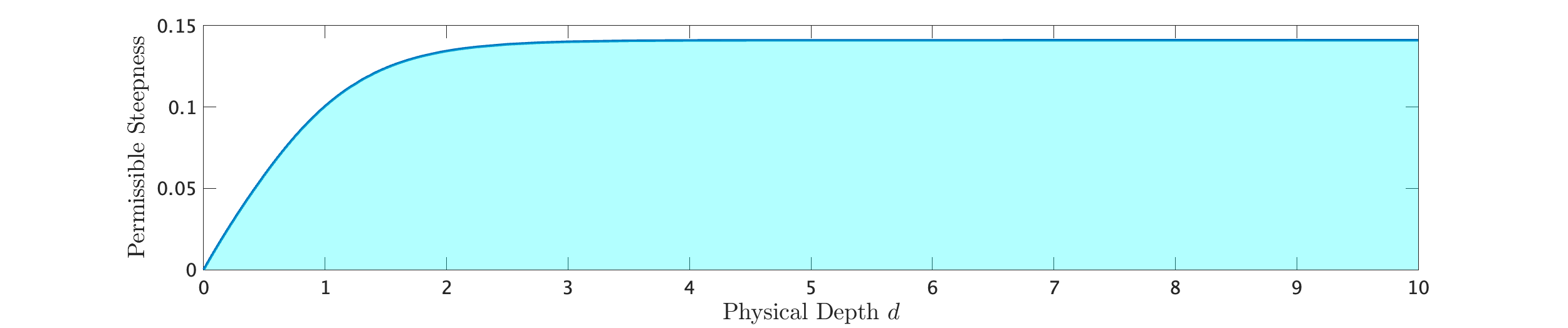}
    \caption{Permissible steepness of Stokes waves as a function of depth. The solid blue line indicates the steepness of the limiting wave in each depth, see Section \ref{sec:HighAmp} for more detail.}
    \label{fig:limitingWaveAmp}
\end{figure}

\no where $g$ is the acceleration due to gravity, $\eta$ is the profile of the free surface and $\phi$ is the velocity potential of the fluid within the bulk: $(\phi_x,\phi_y)$ is the fluid velocity. In 1843, Stokes \cite{stokes1847theory,stokes1880supplement} studied periodic traveling solutions of this problem. The existence of such waves, now called Stokes waves, in infinite depth was proven by Levi-Civita~\cite{levi1925determination} and Nekrasov~\cite{nekrasov1921waves}. Struik~\cite{struik1926determination} generalized their result to water of finite depth. Stokes \cite{stokes1880supplement} conjectured that there exists a wave of maximum amplitude that has a 120$\degree$ corner at the crest of the wave. In 1951, Nekrasov \cite{krasovskii1962theory} found an integral equation describing the slope of the steepest wave near its crest. Both Krasovskii \cite{krasovskii1962theory} and Keady and Norbury \cite{keady_norbury_1978} used this equation to study the limiting wave. Toland \cite{toland1978existence} improved upon their method to prove the existence of the limiting wave in 1978. Independently, Plotnikov \cite{plotnikov1982proof,plotnikov2002proof} proved the same in 1982. In the same year, the conjecture regarding the angle at the limiting wave's crest was proven by Amick \textit{et al}  \cite{amick1982stokes}. Later, Amick and Fraenkel \cite{amick1987behavior} and McLeod\cite{mcleod1987asymptotic} studied the behavior of the steepest wave near the crest using perturbative methods. A recent overview of the history of this problem is found in~\cite{haziot2022traveling}. In this article, we consider finite-depth Stokes waves.  Figure~\ref{fig:limitingWaveAmp} shows the parameter space of permissible amplitudes for a Stokes wave as a function of depth, obtained using our methods. 

Both the speed $c$ and the energy $\mathcal{H}$ of a Stokes wave can by parameterized by its steepness $\mathcal{S}=A/L,$ the ratio of the 
crest-to-trough height of the wave and its wavelength. Preceding Amick \textit{et al}'s \cite{amick1982stokes} proof, Longuet-Higgins and Fox \cite{longuet1977theory,longuet1978theory} used perturbation theory for water of infinite depth to find that the speed and energy undergo at least one oscillation as the steepness is increased and the limiting wave is approached. In fact, their result indicates the presence of an infinite number of oscillations as the limiting values of the speed, energy, etc. are approached. More recently, Dyachenko \textit{et al} \cite{Dyachenko2023Almost} and Deconinck \textit{et al}  \cite{deconinck2023dominant} verified this numerically for the first few oscillations. We are not aware of any extensions of the work of Longuet-Higgins and Fox to waves traveling on a finite-depth fluid.

Other studies on steep Stokes waves in water of finite depth do exist. Cokelet \cite{cokelet1977steep} computed the amplitudes of the waves which extremize both the speed and the energy in a large number of finite depths, as well as the change of these quantities through the computer-aided solution of a Pad\'e-based asymptotic series. More recently, Zhong and Liao \cite{zhong2018limiting} implemented a homotopy analysis method which enables them to go beyond Cokelet's computations, and approximate many quantities related to the limiting wave. Neither of these methods directly solve the equations, instead relying on the convergence of various power series. Furthermore, while Zhong and Liao's work \cite{zhong2018limiting} significantly extends Cokelet's computational domain, their techniques rely on a relatively small number of Fourier modes and their results for the amplitude of the steepest wave in infinite depth are less accurate than (and their significant digits disagree with) those of Dyachenko \textit{et al} \cite{Dyachenko2023Almost}. Since the finite-depth methods presented in this manuscript build on those of \cite{dyachenko2019stokes,Dyachenko2023Almost,dyachenko2014complex,dyachenko2016branch,semenova2024stokes}, we expect them to be more accurate than those in \cite{zhong2018limiting}. 

Using the Pad\'e-based method on which Cokelet's work is based, Schwartz \cite{schwartz1974computer} detected the presence of a square-root type singularity above the crest of a Stokes wave, outside the fluid domain. This agrees with the previous results of Grant \cite{grant1973singularity}, who argued that the Bernoulli condition~\eqref{eq:dynamicEuler} implies that potential singularities above the fluid are necessarily of square-root type. This work was built on by Tanveer~\cite{tanveer1991singularities}, who confirmed the square-root branch type of the singularities and developed numerical algorithms that incorporated this. His work found one singularity above the fluid in both finite and infinite depth and gives an indication that there are an infinite number of singularities on non-physical Riemann sheets connected to those branch points. Twenty years later, a series of papers analyzed the singularity structure of Stokes waves: in 2014, Dyachenko \textit{et al}~\cite{dyachenko2014complex} approximated the location of the singularity above the fluid domain and determined the rate of its approach to the crest of the steepest wave as steepness is increased. Dyachenko \textit{et al} \cite{dyachenko2016branch} refined these methodologies through the use of Pad\'e approximations, which Lushnikov~\cite{lushnikov2016structure} used to investigate the structure of the Riemann sheets associated with this branch point with great accuracy. Crew and Trinh~\cite{crew2016new} investigated this aspect of the singularity structure as well. Both papers provide compelling evidence for the mechanism by which the square-root-type branch point of steep Stokes waves becomes a cube-root type branch point for the limiting wave. None of the works more recent than Tanveer's have been extended to the finite-depth case. We initiate this investigation in Section \ref{sec:Pole Structure}.

The outline of our paper is as follows. In Section \ref{sec:ProblemFormulation}, we describe the conformal mapping reformulation of (\ref{eq:Laplace}-\ref{eq:dynamicEuler}). In Section \ref{sec:NumericalMethod}, the method of Semenova and Byrnes \cite{semenova2024stokes} is extended to the computation of Stokes waves in fixed physical depth. In Section \ref{sec:LHFox}, we extend the results of Longuett-Higgins and Fox~\cite{longuet1977theory,longuet1978theory} to finite depth.
%including comparing the computation of the various bifurcation points to more general wave types in Section \ref{sec:BifAnalysis}. Section \ref{sec:BifAnalysis} also discusses the connections of these bifurcation points to the change of the stability spectra of Stokes waves, as observed in infinite depth~\cite{deconinck2024Recurrence}. 
Finally, in Section \ref{sec:Pole Structure} we discuss the singularity structure of the waves and potential extensions of the results of Dyachenko \textit{et al} \cite{dyachenko2016branch}, Lushnikov \cite{lushnikov2016structure}, and Crew and Trinh \cite{crew2016new} to finite depth.
%develops conjectures relating to the spectra of these waves through a discussion of the Taylor sign condition. 

\section{Problem Formulation}\label{sec:ProblemFormulation}

\begin{figure}
    \centering
    \includegraphics[width=\linewidth]{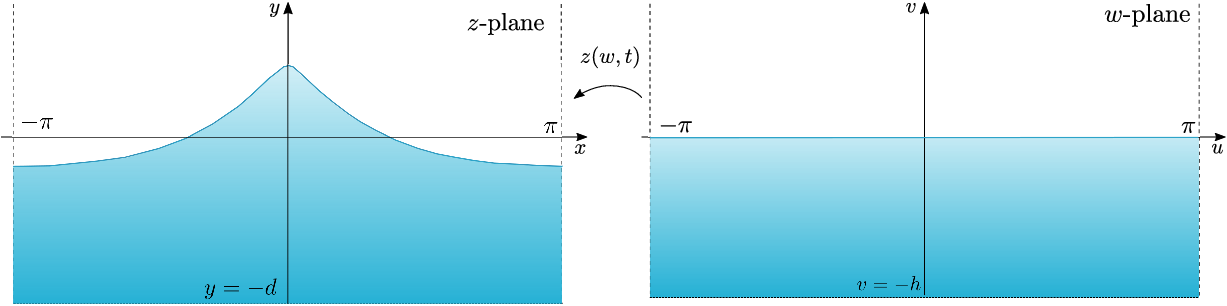}
    \caption{The mapping $z(w,t)$ from the conformal domain (right) to the physical domain (left). The free surface of the physical domain $(x(u,t),y(u,t))$ is parameterized as a function of the top surface of the conformal domain $w = u + i0$.}
    \label{fig:conformalMapping}
\end{figure}

We follow \cite{dyachenko1996analytical,semenova2024stokes} and use the conformal mapping reformulation of the water wave problem. This reformulation views the parameterization of the surface $(x(u,t),y(u,t))$ as the real and imaginary parts of a conformal mapping $z(w)$ from the upper boundary of a rectangular domain $w = u + iv$ of known width $2\pi$ and height $h$ in the complex plane, see~Figure~\ref{fig:conformalMapping}. 
The velocity potential along the surface is defined as $\psi(u,t) = Re [\Pi(u,t)]$, using
the complex potential $\Pi(w,t) = \phi(w,t)+i\theta(w,t)$. %We define the velocity potential along the surface to be $\psi(u,t) = Re [\Pi(u,t)]$.
%The mapping between the conformal and physical domains is shown in Figure~\ref{fig:conformalMapping}. 

Using this reformulation, the four explicit PDEs posed on an evolving domain with a free boundary (\ref{eq:Laplace}-\ref{eq:dynamicEuler}) become the pair of non-local, implicit PDEs posed on the interval $[-\pi,\pi]$ given by

\begin{align}
    y_tx_u-x_ty_u &= - \hat{R}\psi_u, \label{eq:kinematicImplict}\\
 \psi_t y_u - \psi_u y_t &= - gyy_u - \hat{R}(\psi_t x_u - \psi_u x_t + gyx_u).\label{eq:dynamicImplict}
\end{align}

\no Here $\hat{R}$ is a generalization of the periodic Hilbert transform which depends on the height $h$ of the conformal domain:

\begin{align}
\hat{R}f(u) = \frac{1}{2h} \dashint_\mathbb{R}\frac{f(u')}{\sinh(\pi (u'-u)/2h)}du',
\end{align}

\no where $\dashint_\mathbb{R}$ is the principal value integral over the real line. To apply this operator using the Fast Fourier Transform, we use

\begin{align}
(\hat{R}f)_k = i\tanh(kh)f_k, \label{eq:RhatFourier}
\end{align}

%which comes from Zakharov \textit{et al}'s 1996 paper.
%Informally, \ref{eq:RhatFourier} states that the $k$th Fourier mode of the function $\hat{R}f$ is equal to the $k$th Fourier mode of the function $k$ multiplied by $i\tanh(kh)$.

\no where $f_k$ is the $k$-th Fourier mode of $f(u)$, see \cite{zakharov1996dynamics}. The operator $\hat{R}$ has an inverse $\hat{T}$, see \cite{zakharov1996dynamics}, given by 

\begin{align}
    \hat{T}f = \frac{1}{h} \dashint_{\mathbb{R}} \frac{f(u')}{1-e^{-\pi(u-u')/h}}du'.
\end{align}

\no In Fourier space, this operator is applied using

\begin{align}
(\hat{T}f)_k = -i\coth(kh)f_k.\label{eq:ThatFourier}
\end{align}

\no Defining the operator $\hat{K} = \hat{T} \partial_u$ and moving into a traveling frame with speed $c$, we find that the surface wave profile satisfies the Babenko equation

%\begin{align*}
%    y_tx_u-x_ty_u &= - \hat{R}\psi_u, \\
%x_t\psi_u - x_u\psi_t + \hat{T}(y_t\psi_u - y_u\psi_t) &= g\left(x_uY + \frac{1}{2}\tilde{k}Y^2\right).
%\end{align*}

%\no Applying a change of variables to a frame moving with speed $c$, (\ref{eq:kinematicImplict}-\ref{eq:dynamicImplict}) become 

%\begin{align}
%    y_tx_u -y_ux_t  &=-\hat{R}\phi_u,\label{eq:TravelingConformalKinematic}\\
%    \phi_tx_u-\phi_ux_t+\hat{T}[y_u\phi_t-\phi_uy_t] + 2c\hat{T}y_t &= \hat{S}y, \label{eq:TravelingConformalDynamic}
%\end{align}

\begin{align}\label{eq:babenkoOp}
0=\hat{S}y = (c^2\hat{K}-g)y-g\left(y\hat{K}y+\frac{1}{2}\hat{K}y^2\right).
\end{align}

\no Here $\hat{S}$ is the finite-depth generalization of the Babenko operator discussed in \cite{semenova2024stokes}. The surface elevation $y$ is related to the velocity potential $\psi$ through 
\begin{align}cy_u=\hat{R}\psi_u.\label{eq:phitoy}\end{align} Our primary concern is the accurate solution of the Babenko equation $\hat{S}y = 0$ for a fluid of fixed physical depth.

The Hamiltonian of the system (\ref{eq:Laplace}-\ref{eq:dynamicEuler}) is also its total energy,

\begin{equation}
    \mathcal{H} = \frac{1}{2} \int_{-\pi}^{\pi} \psi \hat{G}(\eta)\psi\,dx + \frac{g}{2}\int_{-\pi}^{\pi} \eta^2\,dx,\label{eq:Energy}
\end{equation}

\no where $\hat{G}(\eta)$ is the Dirichlet-to-Neumann operator~\cite{craig1993numerical, zakharov1968stability}. In conformal variables, the Hamiltonian is written as

\begin{equation}
\mathcal{H} = -\frac{1}{2}\int_{-\pi}^{\pi}\psi \hat R \partial_u \psi du +\frac{g}{2}\int_{-\pi}^{\pi}y^2x_udu,\label{eq:Energy_conf_variables}
\end{equation}

\no with Dirichlet-to-Neumann operator $-\hat R \partial_u$,
see~\cite{dyachenko1996analytical} for details.

\vspace*{0.1in}

\no {\bf Remark.} In all computational results and figures that follow, the acceleration of gravity $g$ is equated to $1$, which is justified through the use of a scaling symmetry.

\section{Computing Waves in a Fixed Depth: Numerical Method}\label{sec:NumericalMethod}

Following Dyachenko \textit{et al} \cite{dyachenko2016branch,semenova2024stokes} in infinite depth, we compute the Stokes waves using a Newton-Conjugate-Gradient method. Waves with a fixed conformal depth $h$ are computed in \cite{semenova2024stokes}. This method will not be expounded upon in this paper. Instead, we extend the application of the method in \cite{semenova2024stokes} to the computation of Stokes waves in two ways. 

\begin{itemize}

\item We use the method of \cite{semenova2024stokes} to compute waves of a fixed {\it physical} depth $d$ instead of a fixed conformal depth $h$. Zakharov {\it et al} \cite{zakharov1996dynamics} demonstrate that $h$ and $d$ differ for non-flat surfaces, see also Dyachenko \& Hur \cite{dyachenko2019stokes}. They are related by the average of the wave's surface elevation $y$ through

\beq 
h = d + \langle y \rangle. \label{eq:DepthRel}
\eeq 

\no As in \cite{dyachenko2019stokes}, $\langle \cdot \rangle$ denotes the average over a single period of the conformal variable $u$ instead of over physical space $x$. 
%We now use this result to compute waves for a fixed physical depth. 
%\textcolor{red}{\sout{Suppose we wish to compute a family of waves in water of physical depth $d=D$. For each value of the speed $c$, we want to choose a value of the conformal depth $\tilde{h}(c)>0$ so that (\ref{eq:DepthRel}) gives}}
To compute a family of waves in water of physical depth $d=D$, for each value of the speed $c$, we choose a value of the conformal depth $\tilde{h}(c)>0$ so that (\ref{eq:DepthRel}) gives

\beq
E(c;\tilde{h},D) = D + \langle y(\tilde{h}(c;D),c) \rangle - \tilde{h}(c;D) = 0.\label{eq:DepthErr}
\eeq 

\no Here, $E(c;h,D)$ measures the difference between the desired value of the physical depth $D$ and the actual value of the physical depth when $h$ is chosen as the conformal depth. When $E(c;h,D)=0$, we find the correct value of the conformal depth $h$ to ensure the resulting wave has the desired value of the speed $c$ in physical depth $D$.

To determine the value $h(c;D)$ that gives rise to a wave traveling at speed $c$ in water of depth $D$, we use a bisection algorithm. We compute a wave and with it the value $E(c;h,D)$ at each step, terminating the calculations once sufficient accuracy (we choose $10^{-11}$ in our computations) is attained. Next, we use the value of $h$ found to compute a wave with the desired speed $c$ and physical depth $D$. A plot of a few Stokes waves of varying steepness in depths 0.1, 1, and 10 is shown in Figure~\ref{fig:WavesDemo}. Figure~\ref{fig:D01Zooms} shows the waves in depth 0.1 from Figure~\ref{fig:WavesDemo} in more detail.

Alternatively, one could use Newton's method to solve \eqref{eq:DepthErr}. This requires the computation of $\partial E/\partial h$. To this end, we compute $\partial y/\partial h$, obtained from a derivative of the Babenko equation (\ref{eq:babenkoOp}) with respect to the parameter $h$ and solving the resulting equation for $\partial y/\partial h$. This requires the inversion of the linearization of $\hat{S}$ about $y$, and hence the use of MINRES at each step of Newton's Method. Although Newton's method requires fewer iterations than does bisection, this added use of MINRES  substantially increases the cost of each Newton step. The bisection method described above is more efficient, even more so as steeper waves are computed. 

\begin{figure}
    \centering
    \includegraphics[width=\linewidth]{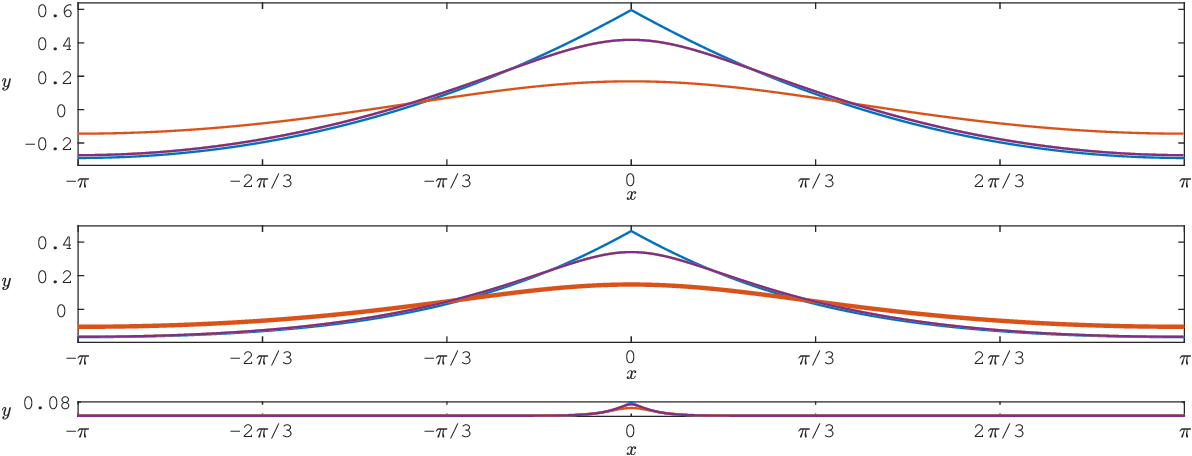}\\

    \caption{Waves of increasing steepness in depths 0.1 (bottom), 1 (middle), and 10 (top). All waves are plotted with an accurate aspect ratio. Note that in each case, a corner with an included angle of 120 degrees forms at the crest. In depth 0.1, the wave localizes to a fraction of the wavelength, reminiscent to a solitary wave. The wave in depth 0.1 is difficult to see at this scale, see Figure~\ref{fig:D01Zooms}.}
    \label{fig:WavesDemo}
\end{figure}

\begin{figure}
    \centering
    \includegraphics[width=\linewidth]{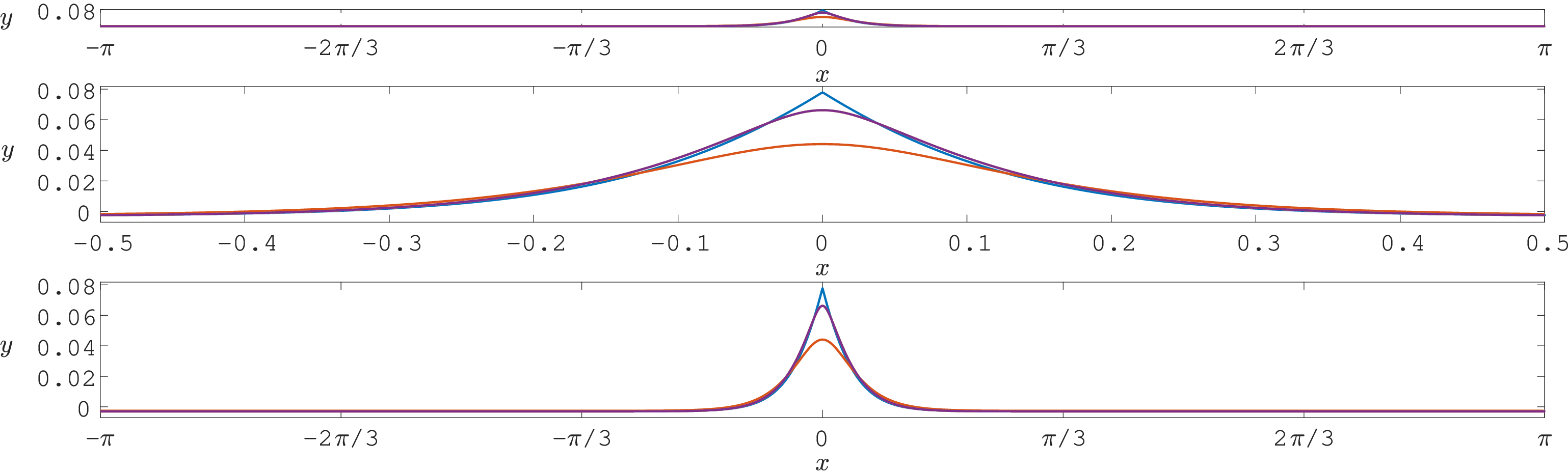}
    \caption{Waves of increasing steepness in depth 0.1. The top two panels are plotted with accurate aspect ratios, with the middle panel being a zoom of the central unit interval of the period to showcase the similarity to the full period of waves in deeper water. The bottom panel has been stretched in the $y$ direction to show the localization of the wave and its profile.}
    \label{fig:D01Zooms}
\end{figure}

As noted in previous works \cite{deconinck2011instability, semenova2024stokes} and shown in Figures \ref{fig:WavesDemo} and \ref{fig:D01Zooms}, in shallow water high-amplitude Stokes waves become increasingly localized within their wavelengths. To measure this localization, we first determine the height of half width of an extreme wave in deep water, as a proportion of the full amplitude of the wave. Then, 
the width of an extreme wave in a shallower fluid at the height which is the same fraction of the full amplitude gives a measure of the localization of that wave. In particular, dividing the full period by double this width gives a modified wave number $\tilde{k}(d)$ for the extreme waves in that depth which accounts for their localization\footnote{To measure the width of half height of an extreme wave in deep water, we use the steepest computed wave in a depth 10 fluid. This serves as a good proxy for an extreme wave in infinite depth as the Fourier signs of the operators in Equation (\ref{eq:babenkoOp}) differ from the infinite depth problem by less than $10^{-6}\%$ at this depth.}. A sample wave in depth 0.5 is shown in Figure~\ref{fig:widthOfHalfHeight}, and the change of the modified wave number as a function of depth is shown in Figure~\ref{fig:modifiedWaveNumber}. Interestingly, this is approximated well by an inverse power of a shifted hyperbolic tangent, also shown in Figure~\ref{fig:modifiedWaveNumber}. Other functions which approach zero linearly as their argument tends to zero and one as their argument tends to infinity could have been used with similar relative errors. However, the hyperbolic tangent offers a connection with the dispersion relation $\omega(k,d)^2=k\tanh(kd)$ with the wavenumber $k$ set to 1. This connection with the dispersion relation is intriguing given the connection between the behavior of the  dispersion relation and the limiting waves in the context of simpler fluid models \cite{ehrnstrom2023precise,enciso2018convexity}. %I'm really not sure about this. Do either of you have opinions? I could email Mats asking for more sources?

\begin{figure}
    \centering
    \includegraphics[width=\linewidth]{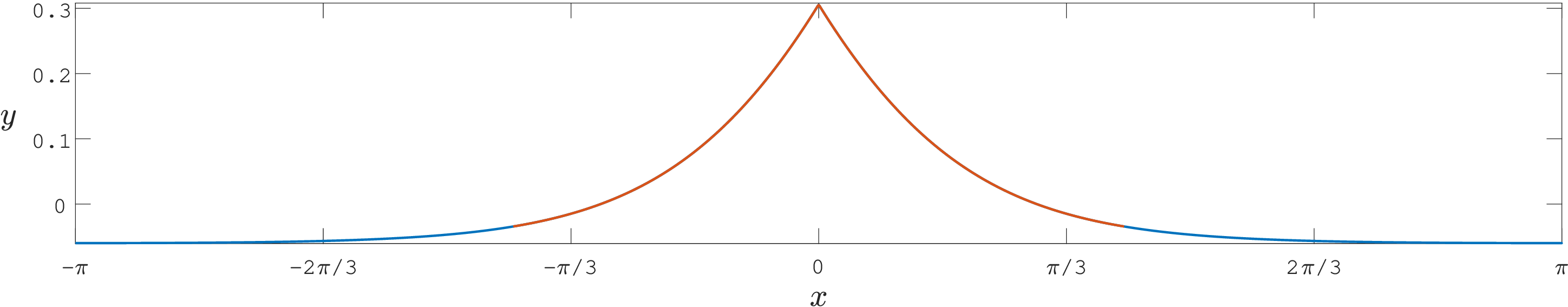}
    \caption{Profile of the steepest computed wave in depth $d = 0.5$. Blue shows the full period of the wave, whereas red shows the localization of the wave within the period, \textit{i.e.} it shows the portion of the wave which is not small. As depth decreases, the width of the red curve shrinks relative to the period as shown in Figures \ref{fig:WavesDemo} and \ref{fig:modifiedWaveNumber}.}
    \label{fig:widthOfHalfHeight}
\end{figure}

\begin{figure}
    \centering
    \includegraphics[width=\linewidth]{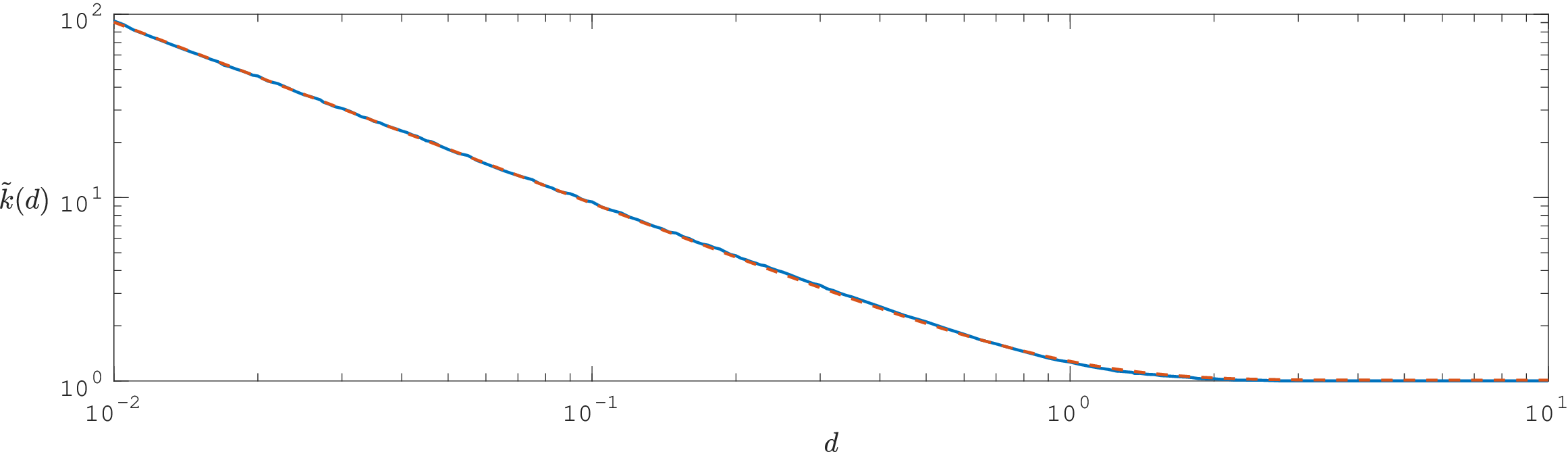}
    \caption{Top: change in the modified wave number $\tilde{k}(d)$, and fit to %a {rescaled} inverse power of $\tanh$,
    $1.06(\tanh(1.01d))^{-0.989}$ where the exponent includes -1 in its confidence interval. The errors in the fit are two orders of magnitude smaller than the value of the modified wave number. Data for the modified wave number is a solid blue line, and the fit is a dashed blue line.}
    \label{fig:modifiedWaveNumber}
\end{figure}

\item A second extension of our work is the execution of these computations in variable precision arithmetic. For these waves, Julia's BigFloat class \cite{bezanson2017julia} is used to solve the Babenko equation $\hat{S}y=0$ with a tolerance of $10^{-64}$, \textit{i.e.}, the computed value of the $2$-norm satisfies \begin{align}\|\hat{S}y\|_2 <10^{-64}.\end{align} %All computed wave files, as well as each wave's associated metrics, are available at *****LINKHERE*****.

Applying this precision increases the costs of the computations significantly and a full exploration of finite-depth Stokes waves with this accuracy will have to wait until the development of a more refined method, discussed briefly at the end of this section as well as in Section \ref{sec:Pole Structure}. Any computations which rely on these waves are mentioned explicitly. All other computations are performed using Matlab in double precision.

\end{itemize}

\begin{figure}
    \centering
    \includegraphics[width=\linewidth]{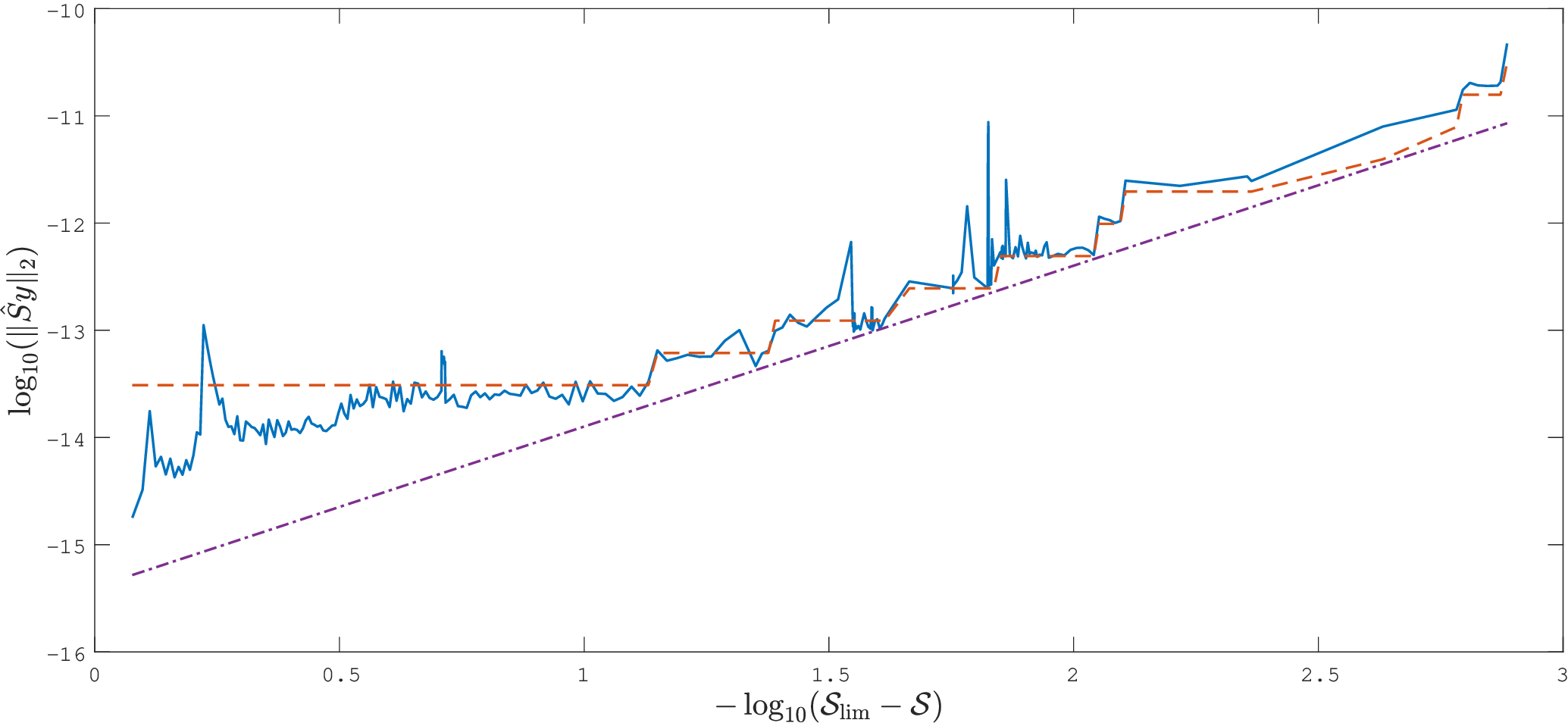}
    \caption{Log-log plot of residuals and number of Fourier modes as a function of steepness. The solid blue line shows the residuals $\|\hat{S}y\|_2$, the dashed red line shows a rescaled number of Fourier modes, and the purple dashed-dotted line is proportional to $(\mathcal{S}_{\lim}-\mathcal{S})^{-1.5}$. The growth of the $L^2$-norm of the residuals is directly related to the number of Fourier modes. Both grow approximately proportionally to $(\mathcal{S}_{\lim}-\mathcal{S})^{-1.5}$. }
    \label{fig:errGrowth}
\end{figure}

%\subsection{Error Analysis}\label{sec:ErrAnalysis}
Since our method is based on a pseudo-spectral method, the application of all operators enjoys spectral accuracy. The convergence of our method is verified by the  residuals of the Babenko equation becoming small, starting around $10^{-14}$ for small waves and eventually reaching errors not exceeding $10^{-10}$ for the steepest waves. This accuracy loss grows proportionately to $(\mathcal{S}_{max}-\mathcal{S})^{-1.5}$ {where $\mathcal{S}_{max}$ is the steepness of the tallest computed wave.} {Such growth} is connected to the approach of a pair of square-root type branch points to the computational domain, discussed in Section \ref{sec:Pole Structure}. In particular, we require an increasing number of Fourier modes to accurately represent the {solution $y$ and the operators in $\hat S$} as this branch point approaches, leading to accumulation of numerical errors. The growth of these errors with the number of Fourier modes as we approach the steepest wave is shown in Figure~\ref{fig:errGrowth}. To compute waves nearer to the limiting wave, one maybe able to follow the work of Lushnikov \textit{et al} \cite{lushnikov2017new} and Dyachenko \textit{et al} \cite{Dyachenko2023Almost} in infinite depth, using an auxiliary conformal mapping to push the singularities further from the conformal domain.

%We also perform a Cauchy convergence study for a number of waves. To do this, we compute a wave with the same values of physical depth $H$ and $c$ increasing $N$ and the highest tolerance allowed by those parameter values. A sample plot, for a wave of steepness slightly greater than that of the first extremizers of the energy in a depth 1 fluid, is shown in Figure~\ref{}. ASK BERNARD WHY THIS IS IMPORTANT (Show residual
%Residuals, Cauchy

\section{Oscillations in Energy and Wave Speed}\label{sec:LHFox}

Longuett-Higgins and Fox \cite{longuet1977theory,longuet1978theory} give asymptotic results relating the profile of a near-limiting wave in infinite depth, as well as its energy and wave speed, to a perturbation parameter $\epsilon$ related to the radius of curvature of that wave. 
{Their perturbation parameter $\epsilon$ is defined as  

\begin{align}
    \epsilon = \frac{s}{c_0\sqrt{2}}, \label{eq:LHEpsilonDefinion}
\end{align}

\no where $s$ is the particle speed at the crest of the traveling wave, and $c_0$ is the phase speed given by the dispersion relation

\begin{align}
    \omega^2(k) = gk\tanh(kd). \label{eq:stokesDispersion}
\end{align}}

\no In particular, Longuett-Higgins and Fox found that the speed of a near-extreme Stokes wave changes as 

\begin{align}
    c^2 &\sim \frac{g}{k}\left(c_{\lim}^2+a_2\epsilon^3\cos(3\mu\ln(\epsilon)+a_4)\right) + \ldots \nonumber\\
    &\sim \frac{g}{k}\left( 1.1931-1.18\epsilon^3\cos(2.143\ln(\epsilon)+2.22)\right)+\ldots .\label{eq:LHSpeed}
\end{align}

\no Here $c_{\lim}$ is the speed of the extreme wave. Similar expansions are given for the kinetic energy $T$ and potential energy $U$:

\begin{align}
    T &\sim T_{\lim}+b_2\epsilon^3\cos(3\mu\ln(\epsilon)+b_4)+\ldots\nonumber\\
    &\sim \frac{g}{k^2}\left(0.03829 -0.215\epsilon^3\cos(2.143\ln\epsilon+1.66)\right)+\dots \label{eq:LHKE},\\
    U &\sim  U_{\lim}+c_2\epsilon^3\cos(3\mu\ln(\epsilon)+c_4)+\ldots\nonumber\\
    &\sim \frac{g}{k^2}\left(0.03457-0.169\epsilon^3\cos(2.143\ln(\epsilon)+1.49)\right) + \dots \label{eq:LHPE}.
\end{align}

\no Longuett-Higgins and Fox \cite{longuet1978theory} also give an expression for the steepness $\mathcal{S}$ as a function of speed, which has a quadratic term in $\epsilon$:
\begin{align}
    \mathcal{S} &\sim \mathcal{S}_{\lim}+ d_1\epsilon^2+ d_2\epsilon^3\cos(3\mu\ln(\epsilon)+d_4) + \ldots \nonumber \\
    &\sim 0.14107-0.50\pi^{-1}\epsilon^2 + 0.160\epsilon^3\cos(2.143\ln(\epsilon)-1.54)+ \ldots.\label{eq:LHSteepness}
\end{align}

These asymptotic results have been verified numerically {by Dyachenko \textit{et al} \cite{Dyachenko2023Almost} and Lushnikov \textit{et al} \cite{lushnikov2017new}, even for waves of steepness below that of the first (\textit{i.e.,} global) extremizer of the total energy. It has been indicated that the energy and speed of a Stokes wave have similar asymptotics in the case of a fixed finite conformal depth \cite{semenova2024stokes}. However, the work of Longuett-Higgins and Fox \cite{longuet1978theory} has not yet been extended to a fixed finite physical depth. Our goal is to give further numerical evidence that these asymptotics extend to finite-depth waves before using these forms to make predictions about the behavior of near-extreme waves in finite depth. The generic asymptotic forms that Longuett-Higgins and Fox derive originate entirely from the kinematic and dynamic surface conditions (\ref{eq:KinematicEuler}-\ref{eq:dynamicEuler}). The only aspect of their calculations that depends on the depth is the calculation of the coefficients in the asymptotic expressions. In particular, they fit all but one of these parameters to the energy and speed of various numerically computed high-amplitude waves in an infinite-depth fluid. The last parameter $\mu$ also depends only on the Bernoulli condition on the free surface. Thus, to find particular asymptotic approximations for the energy, speed, and steepness in a given finite depth, we need only fit the generic forms to the wave data. To do this, we must determine the particular value of the perturbation parameter $\epsilon$ which corresponds to a particular wave. 
%Ellie re-write this
\subsection{Fitting Algorithm}\label{sec:FittingAlgorithm}
%When we solve the Babenko equation, we know the particular value of the wave speed $c$ that our wave travels at, the particular value of the conformal depth $h$, and the physical depth of the water our wave is traveling on. Also, as a result of our calculations we find the value of our surface elevation $y$ and position $x$ as a function of the conformal variable $u$. 
\textbf{Step One. Computing the Perturbation Parameter $\epsilon$} \\
%Longuett-Higgins and Fox \cite{longuet1977theory,longuet1978theory} define their perturbation parameter $\epsilon$ as  
%\begin{align}
%    \epsilon = \frac{s}{c_0\sqrt{2}}, \label{eq:LHEpsilonDefinion}
%\end{align}
%where $s$ is the particle speed at the crest of the traveling wave, and $c_0$ is the phase speed given by the dispersion relation
%\begin{align}
%    \omega^2(k) = gk\tanh(kd). \label{eq:stokesDispersion}
%\end{align}
The particle speed~\eqref{eq:LHEpsilonDefinion} in the moving frame is computed using the velocity potential~$\phi$ using

\begin{align}
s = \left[  \sqrt{(\phi_x-c)^2+\phi_y^2} \right]_{y = \eta}. \label{eq:surfaceSpeed}
\end{align}
%We define the velocity potential on the free surface to be $q_{S}(x,t)=\phi(x,\eta(x,t),t)$. 

\no From \cite{deconinck2011instability}, the derivative of the velocity potential at the surface $\psi$ with respect to $x$ in the moving frame is 

\begin{align}
    \psi_{x}&=c-\sqrt{\left(1+\eta_x^2\right)(c^2-2g\eta)}.\label{eq:qCreedon}
\end{align}

\no Using the chain rule in the moving frame,
$\psi_{x} = \left[\phi_x+\eta_x\phi_y\right]_{y = \eta}.$
From the kinematic boundary condition~(\ref{eq:KinematicEuler}),
\begin{align}-c\eta_x + \left[\eta_x\phi_x -\phi_y\right]_{y = \eta} = 0.\label{eq:contAFMTraveling}\end{align}

\no Using these two equations, one can solve for $\phi_x$ and $\phi_y$ in terms of $\eta_x,$ which itself can be computed as 

\begin{align} \eta_x =y_x= \frac{y_u}{x_u}. \label{eq:etaxchairule}\end{align}

\no The derivatives on the right-hand side of (\ref{eq:etaxchairule}) are computed using a Fast Fourier Transform, guaranteeing spectral accuracy. We see that 

\begin{align}
s = \sqrt{\frac{(\psi_x-c)^2}{1+\eta_x^2}}=\sqrt{c^2-2g\eta},\label{eq:surfSpeed}
\end{align}

\no allowing for the computation of the particle speed from known surface variables directly.\\

\no \textbf{Step Two. Fitting Longuett-Higgins and Fox's Asymptotics}

Having computed the perturbation parameter $\epsilon$ associated with a particular wave, we fit the asymptotic forms given in \cite{longuet1978theory}, repeated in (\ref{eq:LHSpeed}-\ref{eq:LHSteepness}). The resulting fits are used to compute the steepness $S_{\text{lim}}$, speed $c_{\text{lim}}$, and total energy ${\mathcal{H}_{\lim}=} T_{\text{lim}}+U_{\text{lim}}$ of the limiting wave in a particular depth.
We present the values for each of these parameters in a variety of fixed physical depths in Table \ref{tab:LHParams}. 
Other parameters of these fits are reported in Table \ref{tab:otherCoeffs}.
Example fits are shown in Figures \ref{fig:SteepnessFits}, \ref{fig:SpeedFits}, and \ref{fig:EnergyFits}. Figure~\ref{fig:y0Fits} shows the change of the zero mode of the Stokes waves and a fit using these asymptotics. While the zero mode $\hat{y}_0$ was not discussed explicitly in Longuett-Higgins and Fox's original paper \cite{longuet1977theory}, it is connected with the momentum.\footnote{To see this, note that the momentum is given as 
$\mathcal{P} = \int yd\phi$. Furthermore, for a traveling wave $\langle \phi_u\rangle -c\hat{T}y_u = \phi_u$ and $\int ydx = \int yx_udu=\langle y \rangle +\langle y\hat{T}y_u\rangle$. This last quantity is proportional to $\mathcal{P}$.}

The change of these parameters, and in particular, the change of the properties of the limiting wave in shallow water is also of interest. In Figure~\ref{fig:LimitingDecay}, we  plot the change of these limiting quantities going from shallow water to deep water. In addition to being monotonically decreasing for decreasing depth, it is interesting that {the change of} each of these {limiting values as a function of depth} is well approximated by a power of a rescaled version of the modified dispersion relation $\tilde{\omega}(d)=\sqrt{\tilde{k}(d)\tanh(\tilde{k}(d)d)}$. In particular, the limiting speed appears to decay like $1/\tilde{\omega}$, the limiting steepness like $1/\tilde{\omega}^2$, and the limiting energy like $1/\tilde{\omega}^6$. As with the modified wave number $\tilde{k}$, these connections with the dispersion relation are not entirely surprising. In fact, since $\tanh(x)\to1$ as $x\to\infty$ and $\tilde{k}(d)\to\infty$ as $d\to 0$, $\tilde{\omega}(d)$ itself can be approximated well by $\sqrt{\tanh(d)}$, a branch of the dispersion relation with wavenumber 1, which leads to only a small increase in the relative errors of these fits. Furthermore, the steepness of the limiting wave {$\mathcal{S}_{\lim}$} then decays proportional to $\omega^2(1,d)$, the speed {$c_{\lim}$} of that wave proportional to $\omega(1,d)$, and the energy {$\mathcal{H}_{\lim}$} proportional to $\omega^6(1,d).$ Although not shown in Figure~\ref{fig:LimitingDecay}, the limiting value of the zero mode of the surface elevation decays like $\omega^4(1,d)$. Figure \ref{fig:otherCoefsFits} demonstrates that the other coefficients of Longuett-Higgins and Fox's asymptotics decay like shifted scaled integer powers of $\omega(1,d)$ as well.

%Note that the dominant wave number of our Stokes waves is $k=1$, since they are $2\pi$ periodic.
\begin{figure}
    \centering
    \includegraphics[width=\linewidth]{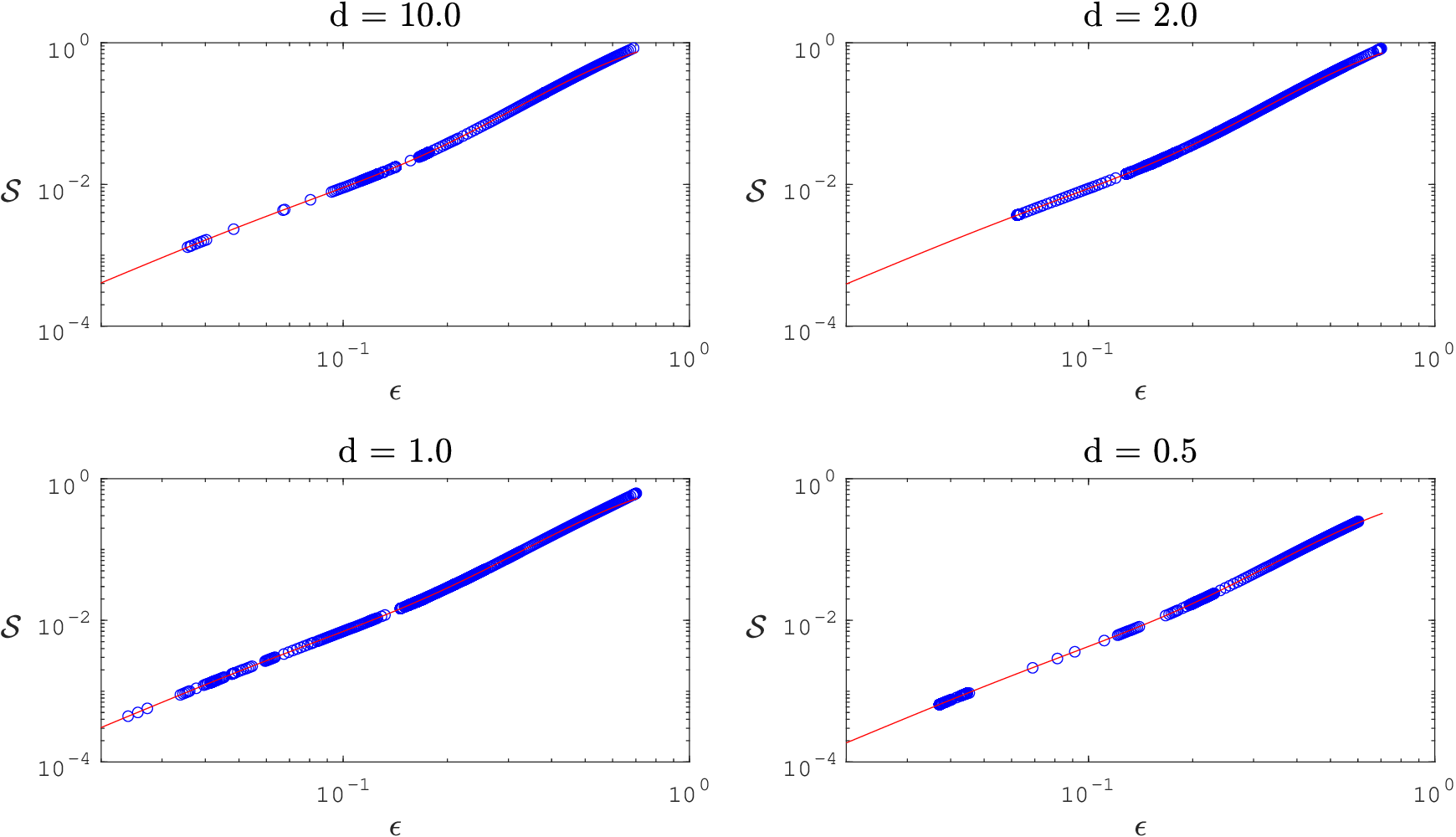}
    \caption{Fit of Longuett-Higgins and Fox's steepness asymptotics (solid red line) to data (blue circles) in depths 10, 2, 1, and 0.5. These fits are more accurate when using a larger number of waves, as in depth 1. The data appears linearly dependent on epsilon with slope 2 for small values of the perturbation parameter $\epsilon$ on the log-log scale, indicating the validity of the asymptotics in finite depth.}
    \label{fig:SteepnessFits}
\end{figure}

\begin{figure}
    \centering
    \includegraphics[width=\linewidth]{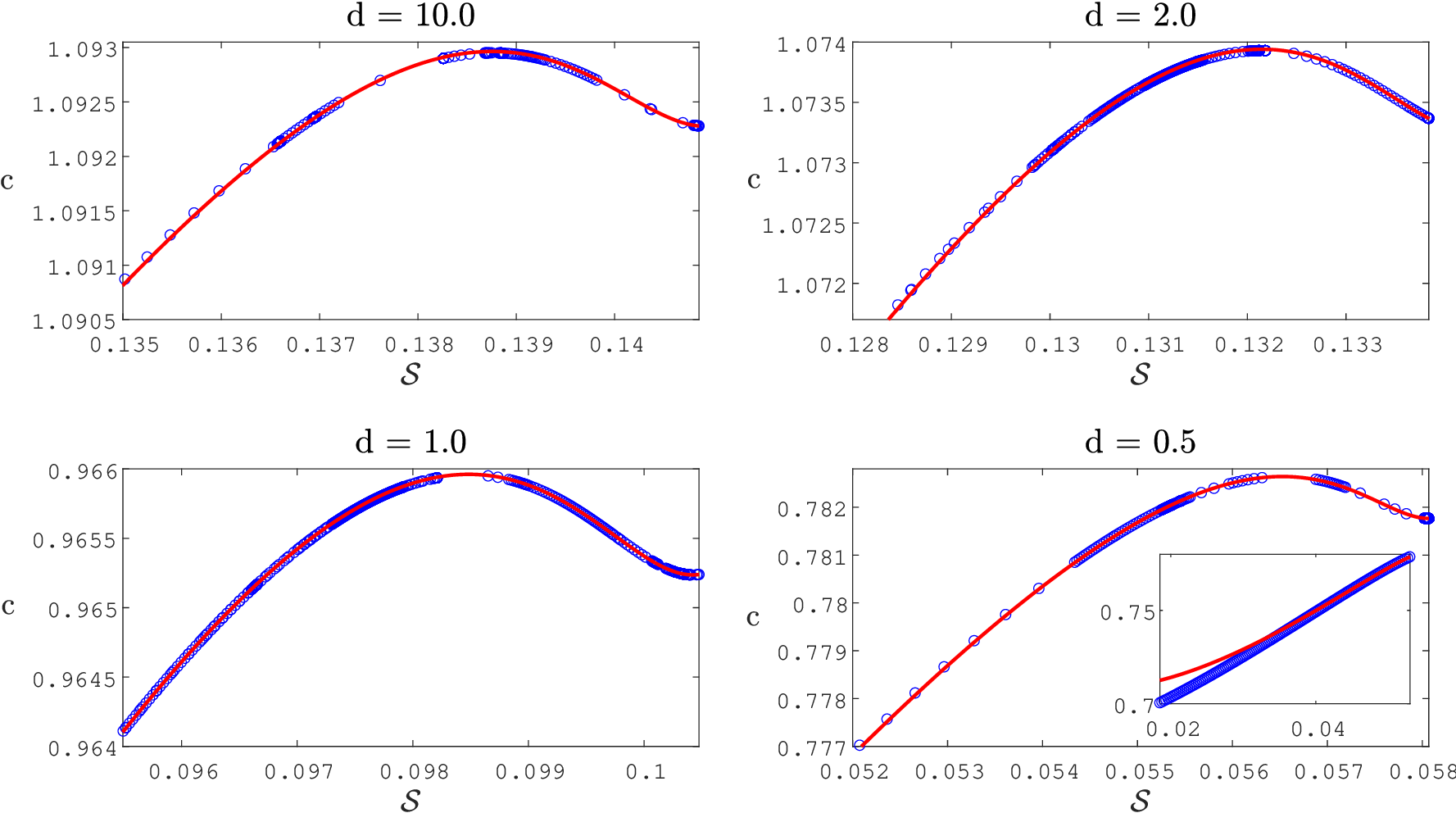}
    \caption{Fit of Longuett-Higgins' speed asymptotics (solid red line) to the data (blue circles) in depths 10, 2, 1, and 0.5. These fits, however good, lose accuracy at small values of steepness. They are more accurate when using a larger number of waves, as in depth 1.}
    \label{fig:SpeedFits}
\end{figure}

\begin{table}
    \centering
    \begin{tabular}{|c||c|c|c|c|}
       \hline
       D  & $\mathcal{S}_{\lim}$ & $\mathcal{S}_{\max}/\mathcal{S}_{\lim}$ & $c_{\lim}$ & $\mathcal{H}_{\lim}$ \\ 
       \hline
       10 & 0.141064(2) & 0.99853(2) & 1.09228(49) & 0.45779(6)\\
       5 & 0.141(10) &0.986(8)& 1.0921(8) & 0.457(8)\\
       2.5&0.13857(6) &0.9951(7) & 1.0851(91) & 0.4397(7) \\
       2 & 0.13442(4) & 0.9956(3) & 1.07326(2) & 0.410(49) \\
       1.75 & 0.13034(79) & 0.99905(5) & 1.061314(5) & 0.383039(5)\\
       1.5 & 0.12399(78) & 0.9979(4) & 1.042199(4) & 0.34199(2) \\
       1.3627 &0.119(28) &0.978(4) & 1.027(20) & 0.312(78)\\
       1.25 & 0.1144(4) &0.97(68)& 1.0119(6) & 0.284(68)\\
       1.00 & 0.100545(4) & 0.99929(7)& 0.965245(3) &0.210998(2) \\
       0.75 & 0.08173(6) &  0.9948(2)& 0.89336(3) & 0.12915(3) \\
       0.5 & 0.058171(2) &0.99822(4)& 0.7817760(5) & 0.055487(3)\\
       0.25&0.03989(4) &0.990(0)&0.59515(5) &  0.0101(4) \\
       0.2 & 0.02504(6) & 0.980(8) &0.5406(56)&0.00561(9)\\
       0.175 & 0.02205(6)& 0.98(79) & 0.50973(6) & 0.003913(4) \\
       0.15 &0.01903(41) &0.9954(7)&0.47568(3) & 0.002563(30) \\
       0.125 & 0.01597(6) & 0.980(5) & 0.4377(69) & 0.00154(59) \\
       0.1&0.012868(90)&0.9977(10)&0.394730(3)&$8.233(62)\cdot 10^{-4}$\\
       0.075 &0.009721(4) & 0.99(39) & 0.3446(80) & $3.6196 \cdot 10^{-4}$ \\
       0.05 & 0.0065(32) & 0.97(2) & 0.2838(0) & $1.1(199)\cdot 10^{-4}$ \\
       0.025 & 0.0032(89)&0.97(1) & 0.2023(83) & $1.4(60)\cdot 10^{-5}$ \\
       0.02 & 0.00263(57) & 0.988(8) & 0.1813(19) & $7.52(6) \cdot 10^{-5}$ \\
       0.0175 & 0.0023(09) & 0.96(8) & 0.1697(56) & $5.07(5)\cdot 10^{-6}$ \\
       0.015 & 0.0019(80)& 0.97(29) & 0.15730(3) & $3.20(7)\cdot 10^{-7}$ \\
       0.0125 & 0.00165(0)& 0.9(700) & 0.14372(1) & $1.86(46) \cdot 10^{-6}$ \\
       0.01 & 0.00132(19)&  0.990(1) & 0.1286(50) & $9.56(5)\cdot 10^{-7}$\\
       
       \hline
    \end{tabular}
    \caption{Limiting parameter values for various depths. Parentheses denote digits which change over a $95\%$ confidence interval provided using Matlab's fit function. The last digit shown in the parentheses is the order of magnitude of the width of the confidence interval, so that $0.01903(41)$ represents that Matlab's $95\%$ confidence is contained in the interval $(0.0190336,0.0190346)$. The first column shows the depth $D$. The second, fourth, and fifth columns show the values of the steepness $\mathcal{S}_{\lim}$, speed $c_{\lim}$, and total energy $\mathcal{H}_{\lim}=T_{\lim}+P_{\lim}$ that the fit predicts for the steepest wave in each given depth. As Longuett-Higgins and Fox's asymptotic forms are more accurate closer to the steepest wave, the third column provides an indication of the accuracy of these predictions.}
    \label{tab:LHParams}
\end{table}

\begin{sidewaystable}[]
    \centering
    \begin{tabular}{|c||c|c|c|c|c|c|c|c|c|c|c|c|c|c|}
    \hline
       D & $a_2$ & $a_4$ & $b_2$ & $b_4$ & $c_2$ & $c_4$ & $d_1$ & $d_2$ & $d_4$\\
       \hline
    10  &-1.19(5)&2.2(01)&-1.2(41)&1.57(6)&-1.06(2)&1.4(50)&-0.99(9)&-1.0(18)&1.5(79)\\
    5 & -1.20(4)&2.1(69)&-1.2(19)&1.5(91)&-1.04(4) & 1.4(59)&-1.0(28)&-1.0(16)&1.4(6)\\
    2.5 & -1.1(83)&2.17(5)&-1.17(4)&1.5(78)&-1.00(4)&1.4(48) & -0.99(3) & -(1.000) & 1.5(3)\\
    2 & -1.14(2) & 2.18(2) & -1.10(1) & 1.56(2) & -0.93(8) & 1.43(3) & -0.96(6) & -0.9(46) & 1.5(4)\\
    1.75 & -1.1(20) & 2.15(5) & -1.01(1)& 1.55(7) & -0.86(1) & 1.42(5) &  -0.93(5)& -0.(89)&1.5(8)\\
    1.5 & -1.0(59)&2.15(4)&-0.91(5)&1.5(31)&-0.77(43)&1.40(1)&-0.89(8)&-0.8(08)&1.5(9)\\
    1.3627 & -1.0(49) & 2.0(85)&-0.79(7)&1.5(41)&-0.675&1.40(2)&-0.(90)&-0.7(7)&1.(42)\\
    1.25 &  -1.01(4)&2.0(7)&-0.72(4)&1.5(22)&-0.6(11)&1.38(4)&-0.8(6)&-0.7(4)&1.4(5)\\
    1.00 & -0.90(6)&2.05(3)&-0.55(2)&1.46(2)&-0.46(0)&1.33(2)&-0.756(3)&-0.6(08)&1.5(51)\\
    0.75 & -0.78(4)&1.9(40)&-0.3(24)&1.3(70)&-0.2(69)&1.2(42)&-0.63(2)&-0.4(90)&1.(493)\\
    0.5 & -0.59(5)&1.76(8)&-0.14(05)&1.26(6)&-0.11(49)&1.1(48)&-0.46(0)&-0.3(50)&1.4(4)\\
    0.25  &-0.34(5)&1.5(20)&-0.024(8)&1.11(3)&-0.0(199)&1.0(09)&-0.24(2)&-0.19(1)&1.3(8)\\
    0.20 &-0.28(5)&1.4(58) & -0.013(3)&1.09(2)& -0.010(70) & 0.9(89)&-0.19(8) & -0.15(5) & 1.(30)\\
    0.175 & -0.25(40) & 1.4(29) & $-0.009(4)$ & 1.0(55) & $-0.007(6)$ & 0.95(7) & -0.17(1) & -0.13(7) & 1.3(3)\\
    0.15 & -0.22(6) & 1.3(90) & $-0.006(61)$ & 1.04(7) & $-0.005(27)$ & 0.95(6) & -0.147(3) & -0.11(8) & 1.3(3)\\
    0.125 & -0.18(72) & 1.3(70) & $-0.003(68)$& 1.0(21) & $-0.0029(3)$ & 0.9(26) & -0.12(4) & -0.09(9) & 1.(28) \\
    0.1  &-0.158(2)&1.33(6)&$-0.0021(1)$&1.0(06)&$-0.0016(79)$ & 0.9(20)&-0.098(61)&-0.08(05)&1.3(03)\\
    0.075 & -0.11(84) & 1.29(5) & $-(8.9)\cdot 10^{-4}$ & 0.9(6) & $-7.(13)\cdot 10^{-4}$ & 0.8(81) & -0.073(7) & -0.060(4) & 1.(30)\\
    0.05 & -0.07(90) & 1.27(4) & $-2.(60)\cdot 10^{-4}$ & 0.9(6) & $-2.0(6)\cdot 10^{-4}$ & 0.8(78) & -0.0(50) & -0.0(40) & 1.(2)\\
    0.025 & -0.040(3) & 1.2(36) & $-3.3(4)\cdot 10^{-5}$ &  0.9(5) & $-2.6(5)\cdot 10^{-5}$ & 0.8(6) & -0.02(47) &  -0.020(6) & 1.(2)\\
    0.020 & -0.032(6) & 1.22(3) & $-1.(80)\cdot 10^{-5}$ & 0.9(2) & $-1.4(3)\cdot 10^{-5}$ & 0.8(3) & -0.019(8) & -0.016(5) & 1.2(3)\\
    0.0175 & -0.028(5) & 1.2(22) & $-1.1(5)\cdot 10^{-5}$ & 0.9(4) & $-9.(1)\cdot 10^{-6}$ & 0.85(9) & -0.017(5) & -0.014(3) & 1.(18)\\
    0.015 & -0.024(3) & 1.22(5) & $-7.(4)\cdot 10^{-6}$ & 0.9(3) & $-5.(87)\cdot 10^{-6}$ & 0.8(47) & -0.01(49) & -0.012(2) & 1.(20) \\
    0.0125 & -0.020(32) & 1.22(3) & $-4.(30)\cdot 10^{-6}$ & 0.93(4) &$-3.(40)\cdot 10^{-6}$ & 0.8(48) & -0.012(3) &  -0.010(3) & 1.(22)\\
    0.01 & -0.016(6) & 1.2(09) & $-2.3(7)\cdot 10^{-6}$ & 0.9(21) & $-1.8(8)\cdot 10^{-6}$ & 0.8(40) & -0.009(8) & -0.008(1) & 1.2(4)\\
    \hline
    
    \end{tabular}
    \caption{Coefficients of the fits of Longuett-Higgins and Fox's asymptotics in various depths. The last digit shown shows the width of the 95\% confidence interval, and parentheses show the digits which change across this confidence interval.}
    \label{tab:otherCoeffs}
\end{sidewaystable}

\begin{figure}
    \centering
    \includegraphics[width=\linewidth]{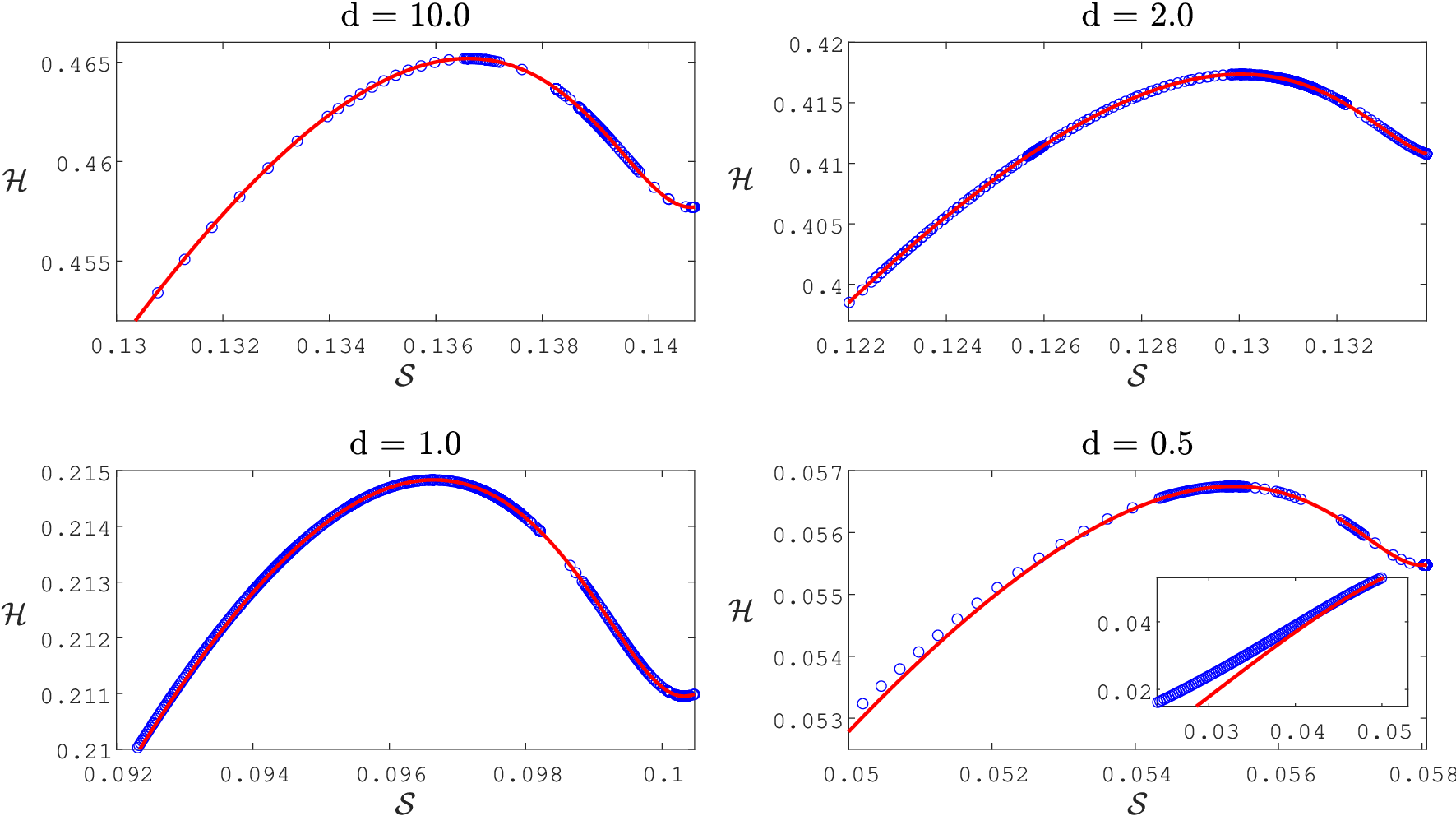}
    \caption{Fit of Longuett-Higgins' energy asymptotics (solid red line) to the data (blue circles) in depths 10, 2, 1, and 0.5. These fits, however good, lose accuracy at small values of steepness. They are more accurate when using a larger number of waves, as in depth 1.}
    \label{fig:EnergyFits}
\end{figure}

\begin{figure}
    \centering
    \includegraphics[width=\linewidth]{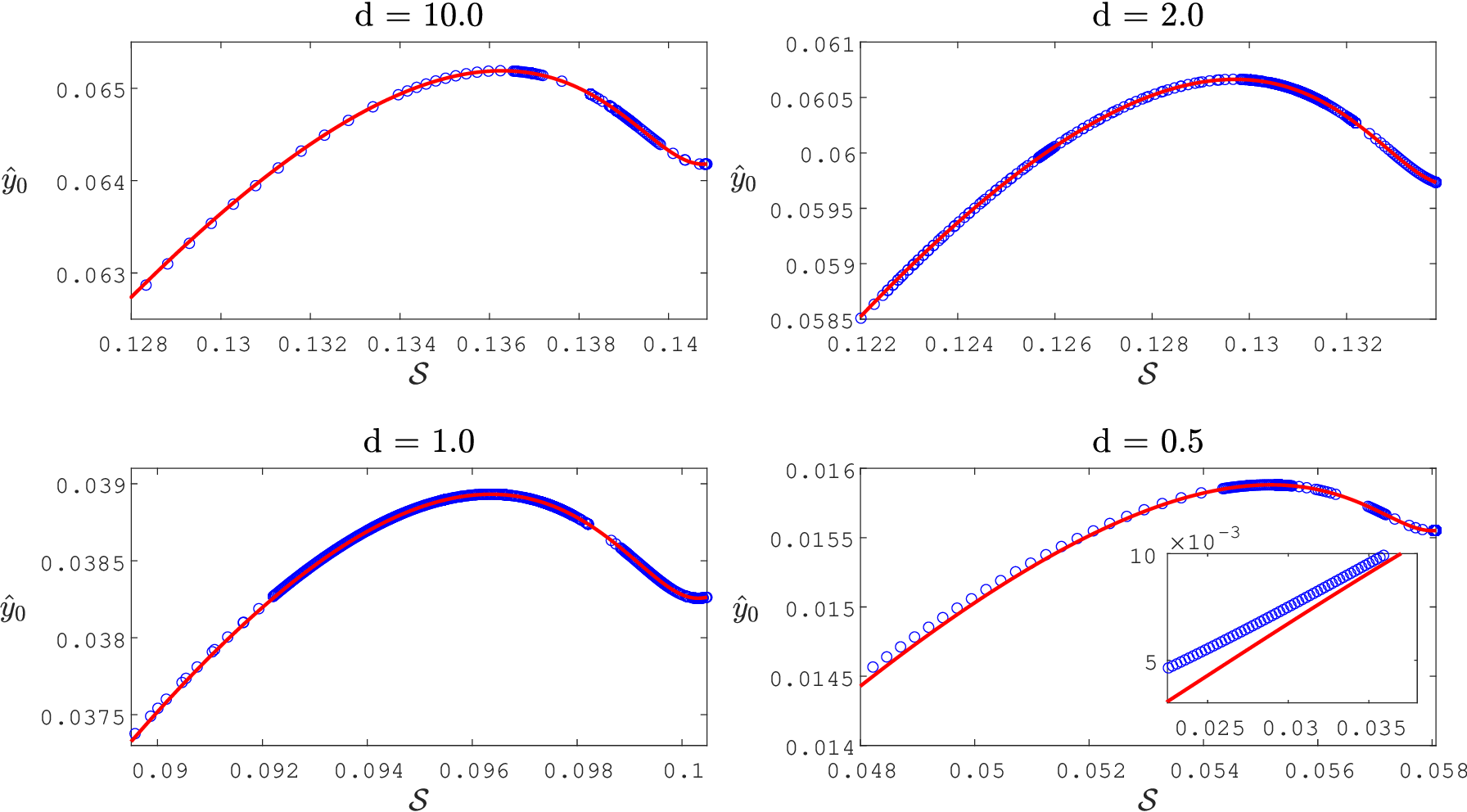}
    \caption{Fit of Longuett-Higgins' asymptotics (solid red line) to the zero Fourier mode of a wave profile $\hat{y}_0=\langle y\rangle$ (blue circles) in depths 10, 2, 1, and 0.5. These particular asymptotics have not been produced in infinite depth. The limiting value of $\langle y\rangle$ appears to decay proportional to $\tilde{\omega}(d)^{-2}$ in the same way as the limiting speed, steepness, and energy are related to inverse powers of the modified dispersion relation. %Since $\hat{y}_0=\langle y \rangle$ relates the physical and conformal depths examined here, however due to it's lack of other physical relevance we will not explore it further.
    }
    \label{fig:y0Fits}
\end{figure}

\begin{figure}
    \centering
    \includegraphics[width=\linewidth]{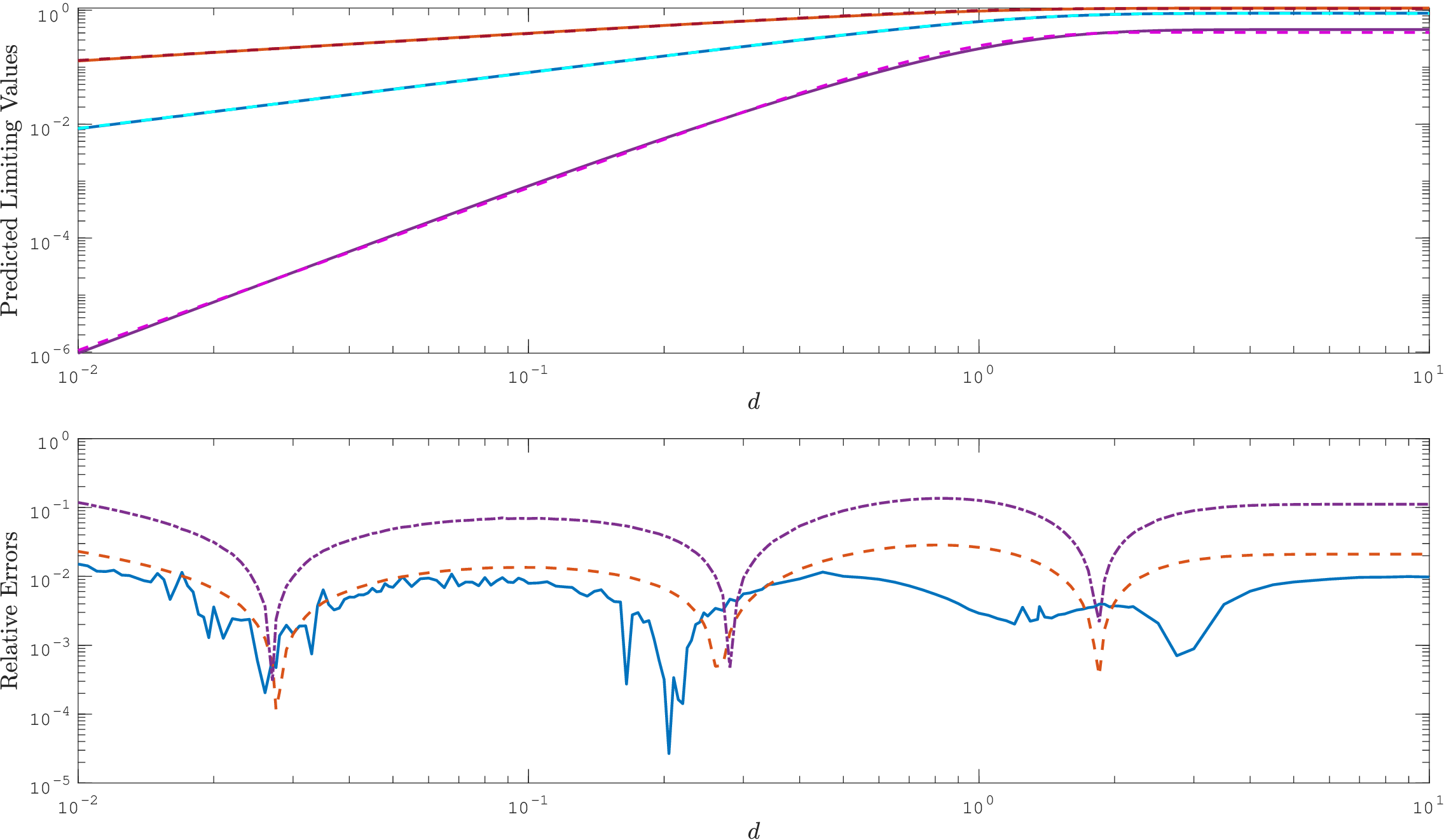}
    \caption{Top: Log-log plot of the predicted energy, speed, and steepness of the limiting Stokes wave as a function of depth, as well as a number of fits to rescaled powers of the modified dispersion relation $\tilde{\omega}(d)=\omega(d,\tilde{k}(d))=\sqrt{g\tilde{k}(d)\tanh(\tilde{k}(d)d)}$. In particular, the speed is compared to $1.074\tilde{\omega}(0.81d)^{-0.479\cdot 2}$, the steepness to $0.886\tilde{\omega}(0.600d)^{-0.995\cdot 2}$, and the energy to $0.41\tilde{\omega}(0.77d)^{-2.9\cdot 2}.$ %This uses the modified wave number discussed in Section \ref{sec:NumericalMethod}. 
    Data is shown in solid lines, the fit as a dashed lines. The top lines (solid orange and dashed red) correspond to the speed of the limiting wave, the middle lines (solid blue and dashed cyan) correspond to the steepness of the limiting wave, and the bottom pair of lines (solid purple and dashed magenta) correspond to the energy of the limiting wave. Bottom: relative errors between the fit and data. All errors are an order of magnitude smaller than the data. The errors appear structured, indicating that a power of the modified dispersion relation may be the first term in an asymptotic expansion involving these parameters. The solid blue line shows the relative residuals of the steepness fit, the dashed red line shows the relative residual of the speed fit, and the dashed-dotted purple line the relative residual of the energy fit. Figure \ref{fig:otherCoefsFits} shows the decay of other coefficients of (\ref{eq:LHSpeed}).}
    \label{fig:LimitingDecay}
\end{figure}

\begin{figure}
    \centering
    \includegraphics[width=\linewidth]{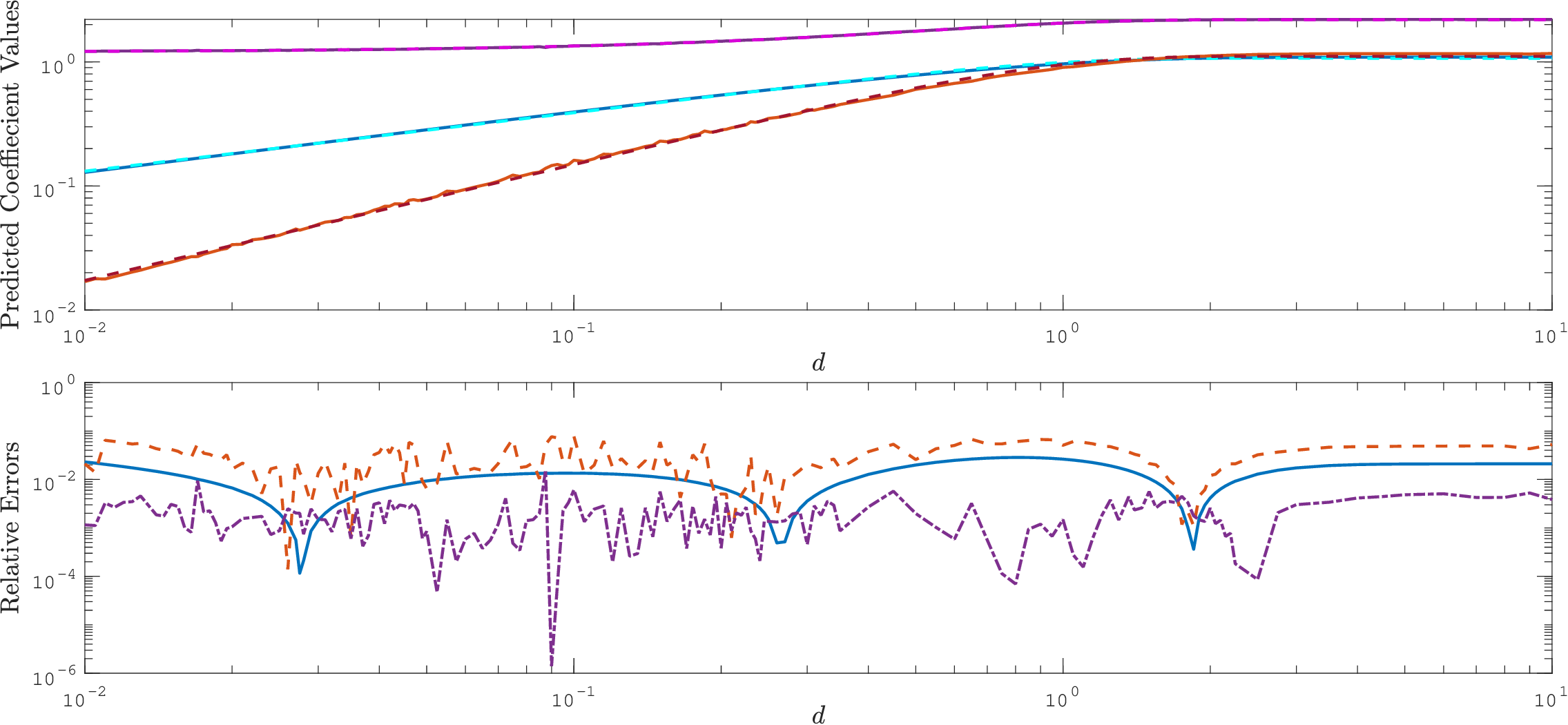}
    \caption{Top: Log-log plot of the coefficients $c_{\lim}$ (Blue), $-a_2$ (red), and $a_3$ (purple) of Longuett-Higgins and Fox's speed asymptotics (\ref{eq:LHSpeed}) as a function of depth, as well as a number of fits (dashed) to rescaled powers of the modified dispersion relation $\tilde{\omega}(d)=\omega(d,\tilde{k}(d))=\sqrt{g\tilde{k}(d)\tanh(\tilde{k}(d)d)}$, with a constant $1.20$ added into the fit of the phase shift $a_3$. In particular, the limiting speed fit is $1.074\tilde{\omega}(0.81d)^{-0.479\cdot 2}$, the $a_2$ fit is $1.1 \tilde{\omega}(0.8d)^{-0.95\cdot 2}$, and the $a_3$ fit is
    $1.19+1.0 \tilde{\omega}(0.8d)^{-0.91\cdot 2}$.
    Bottom: relative residuals for  $c_{\lim}$ (solid blue), $a_2$ (dashed red), and $a_3$ (dash-dotted purple). All errors in the fits are all an order of magnitude less than the data and relatively structured (without the noise from our original fits), indicating this may be the first term in an asymptotic expansion of these coefficients. Similar fits hold for the coefficients of $T$, $U$, $\mathcal{S}$, and $\hat{y}_0=\langle y \rangle$.}
    \label{fig:otherCoefsFits}
\end{figure}

\subsection{Local Extremizers of the Speed and Energy in Varying Depth}\label{sec:HighAmp}

In Section \ref{sec:FittingAlgorithm}, we used the computed Stokes waves to find asymptotic forms for the speed, steepness, and energy as the limiting wave is approached in finite depth. We extract other information from these forms beyond these predicted limiting values. In particular, assuming that we can differentiate the asymptotic relations (\ref{eq:LHSpeed}-\ref{eq:LHSteepness}), we can approximate the steepness of each of the various extremizers of the speed and energy and the extreme values of these quantities by finding the extrema of their asymptotics. 

%The first few of these may be compared with the known values of these extremizers to justify this approach, as discussed below.
%However, we can also extract the steepnesses at which these asymptotic formulas predict various extrema of the speed and energy occur as well as the predicted extremizing values of these quantities.
To find the values of $\epsilon$ that extremize the asymptotic formula for speed, we solve

\begin{align}
    0&=\frac{dc_{LHF}^2}{d\epsilon}=\frac{g}{k}\left[3a_2\epsilon^2\cos(3\mu\ln(\epsilon)+a_4)+a_2\epsilon^3\sin(3\mu\ln(\epsilon)+a_4)\frac{3\mu}{\epsilon}\right],
\end{align}

\no where $c_{LHF}$ is the speed from the Longuet-Higgins and Fox asymptotic relations.
Recall that $a_2$ and $a_4$ are fit parameters from the asymptotic formula (\ref{eq:LHSpeed}) and $\mu$ originates from the surface conditions and is a constant independent of depth. It follows that

\begin{align}
    \epsilon_k=\exp\left(\frac{\left(\pi k+\arctan\left(\frac{1}{\mu}\right)\right)-a_4}{3\mu}\right), \quad k\in \mathbb{Z}. \label{eq:epsSpeed}
\end{align}

\no In (\ref{eq:epsSpeed}), each integer $k$ corresponds to a distinct extremizer of the asymptotic formula (\ref{eq:LHSpeed}). However, only for sufficiently small $k$ in (\ref{eq:epsSpeed}) will $\epsilon$ correspond to values of (\ref{eq:LHSpeed}) which approximate true extreme values of the speed. In particular, if we define $\epsilon_{c,n}$ to be the value of $\epsilon_k$ which approximates the perturbation parameter at the $n$th extremizer of the speed, we have that $n\to\infty$ as $k\to-\infty$. Regardless, it is interesting to note that (\ref{eq:epsSpeed}) depends only on the phase-shift $a_4$, as $\mu$ is constant across all depths. As $c_{LHF}$ is positive for all $\epsilon$, it has the same extremizing $\epsilon$ as $c_{LHF}^2$.

Extremizing the energy is more complicated, as there are two terms. Using sine and cosine addition identities, we solve for the extremizing $\epsilon$. Using the coefficients listed in (\ref{eq:LHKE}-\ref{eq:LHPE}), the extremizers are

\begin{align}
    \epsilon_k = \exp\left(\frac{1}{3\mu}\left[k\pi+\arctan\left(-\frac{[b_2\cos(b_4)+\mu b_2\sin(b_4)+c_2\cos(c_4)+c_2\sin(c_4)]}{[b_2\mu\cos(b_4)-b_2\sin(b_4)+c_2\mu\cos(c_4)-c_2\sin(c_4)]}\right) \right]\right), \quad k\in \mathbb{Z}.\label{eq:epsEnergy}
\end{align}

\no If the denominator of the argument of the arctan is zero, we choose the arctan term to be $\pi$. As in (\ref{eq:epsSpeed}), $k$ parameterizes the different extremizers, with smaller $k$ corresponding to steeper extremizers of the energy. We define $\epsilon_{\mathcal{H},n}$ to be the value of $\epsilon_k$ which approximates the perurbation parameter at the $n$th extremizer of the energy. Since the speed at the crest varies from the phase speed at zero amplitude to 0 at the steepest wave, we are searching for $\epsilon$ in the range $[0,1/\sqrt{2}]$. We plot the evolution of the steepness of the first few extremizers of the energy and the speed in Figure~\ref{fig:EnergySpeedDecay}, and list their steepness and values in Table \ref{tab:ExtremizerLocsVals}. Despite their investigation of extreme waves and the limiting wave in various depths, Cokelet \cite{cokelet1977steep} and others \cite{schwartz1974computer,zhong2018limiting} did not discuss the local extremizers of these quantities. This is a notable gap in the literature considering the connections between these values and interesting points in the change of the topology of the stability spectra \cite{longuet1984stability,longuet1986bifurcation,tanaka1983stability,tanaka1985stability,saffman1985superharmonic,zufiria1986superharmonic,dyachenko2025bifurcations}.

\begin{table}
    \centering
    \begin{tabular}{|c||c|c|c|c|c|c|c|c|}
    \hline
      D & $\mathcal{S}_{\mathcal{H},1}$ & $\mathcal{S}_{c,1}$ & $\mathcal{S}_{\mathcal{H},2}$ & $\mathcal{S}_{c,2}$ & $\mathcal{S}_{\mathcal{H},3}$ & $\mathcal{S}_{c,3}$ &$\mathcal{S}_{\mathcal{H},4}$ & $\mathcal{S}_{c,4}$
      \\ \hline
      10 & 0.858 & 0.872& 0.8846 & 0.8854 & 0.88624 & 0.88628 & 0.88632 & 0.88633 \\
      5 & 0.858 & 0.871 & 0.8848 & 0.8856 & 0.88652 & 0.88656 & 0.886614 & 0.886616 \\
      2 & 0.816 & 0.830 & 0.8429& 0.8437 & 0.84452 & 0.84457 & 0.844611 & 0.844613\\
      1.5 & 0.751 & 0.765 & 0.7775&0.7782&0.77901&0.77905& 0.779097 & 0.779099\\
      1 &0.607 &0.618&0.6303&0.6309&0.6316&0.6317&0.631741&0.631743 \\
      0.5 & 0.347 & 0.355 & 0.3644 & 0.3648 & 0.36544 & 0.36546  &0.365497&0.365498\\
      0.1 & 0.075 & 0.077 & 0.0805 & 0.0806 & 0.080843 & 0.080847 & 0.080856 & 0.080857 \\
      0.01 & 0.0077 & 0.0079 & 0.00827 & 0.00828 & 0.0.0083043 & 0.0083047 & 0.00830583 & 0.00830585\\ \hline 
      D & $\mathcal{H}_1$ & $c_1$ & $\mathcal{H}_2$ & $c_2$ & $\mathcal{H}_3$ & $c_3$ & $\mathcal{H}_4$ & $c_4$\\
      \hline
      10 & 0.4651 & 1.0929 &  0.45770 &1.09227 & 0.457797&1.0922852& 0.457796&1.0922851\\
      5 & 0.4650 & 1.0929& 0.4576 &1.09223 & 0.45771 &1.0922411 & 0.457709 & 1.09224108 \\
      2 & 0.4173 &1.0739 &0.4105 &1.07325 & 0.410643 &1.0732657 & 0.410642&1.0732656\\
      1.5 & 0.3477& 1.0428 & 0.34192 & 1.042191 & 0.341992& 1.0421999 &0.341991&1.0421998 \\
      1 & 0.2148& 0.9659 & 0.21095& 0.96523 & 0.2109989& 0.9652454 & 0.2109983 & 0.9652453\\
      0.5 & 0.0567 &0.7826&0.05547 &0.78176& 0.0554877 &0.7817762&0.0554875 &0.7817761 \\
      0.1 &0.0008491& 0.3955& 0.0008230& 0.39471&0.000823356&0.3947300&0.000823353&0.3947299\\ \hline 
      
    \end{tabular}
    \caption{Predicted steepness of the first 4 extremizers of the speed and energy, and the corresponding values of speed and energy in a number of depths, computed using Longuett-Higgins and Fox's asymptotics.}
    \label{tab:ExtremizerLocsVals}
\end{table}

\begin{figure}[tb]
    \centering
    \includegraphics[width=\linewidth]{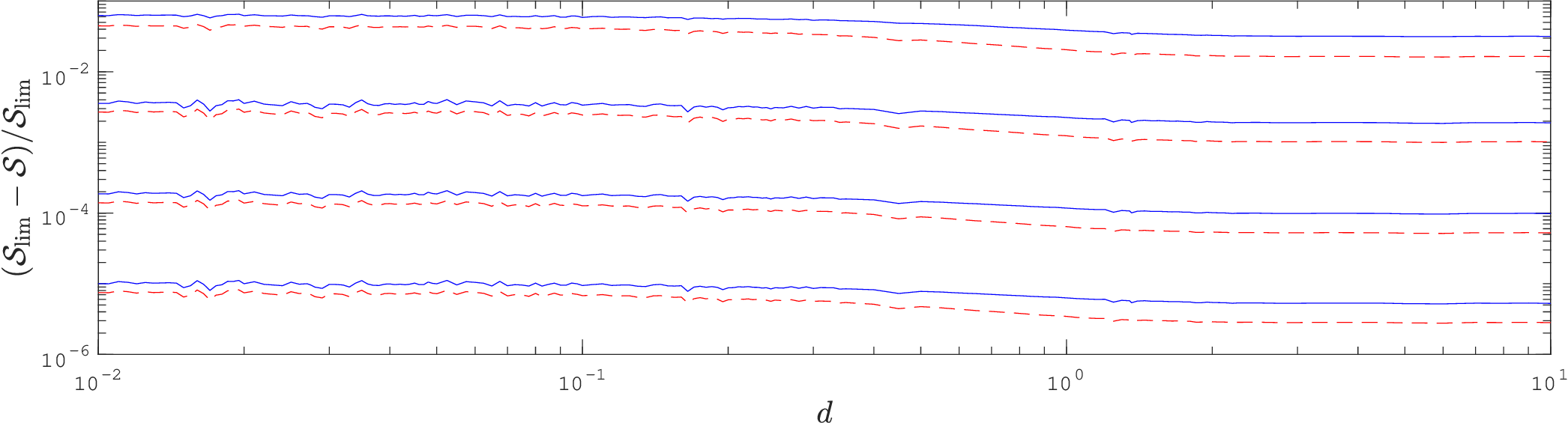}
    \caption{Plot of the predicted steepness of the first few extremizers of the energy (solid blue) and speed (dashed red) relative to the predicted steepness of the steepest wave. Note that these extremizers interlace, as predicted in infinite depth.}
    \label{fig:EnergySpeedDecay}
\end{figure}

We check these results in different ways. First, we compare with a direct computation of the steepness of the extremizers of the speed and energy. To do this, we find the solutions to (\ref{eq:babenkoOp}) which extremize the speed or energy. %Further, we find the steepness of waves which extremize the speed by looking for zero eigenvalues of the linearized Babenko operator $\hat{S}_1(y)$, as discussed in Section \ref{sec:BifAnalysis}. 
This comparison is found in Table \ref{tab:ExtremaComparisons}. In addition to knowing the steepness of the various extremizers of the energy and the speed, and how their values change as the limiting wave is approached, we consider a ratio of the steepnesses at which various extremizers occur. In particular, we define $S_{c,n}$, $S_{\mathcal{H},n}$ to be the steepness of the $n$th non-trivial extremizer of the speed and energy, respectively. We are interested in the limit as $n\rightarrow \infty$ of

\begin{table}[tb]
    \centering
    \begin{tabular}{|c||c|c|c|c|c|c|}\hline
        $D$ & $\mathcal{S}_{dir,1}$& $\mathcal{S}_{LHF,1}$ &  $\Delta \mathcal{S}_{1}$& $\mathcal{H}_{dir,1}$  &$\mathcal{H}_{LHF,1}$&$\Delta \mathcal{H}_{1}$\\
        \hline
        10 & 0.858305 &0.85831& $5\cdot 10^{-6}$ & 0.465177 &0.46518 &$3\cdot 10^{-6}$  \\
        
        1  & 0.607248 &0.607278 &$3\cdot 10^{-5}$ &0.214835& 0.214834 & $1\cdot 10^{-6}$\\
        0.1&0.0760123&0.076114  & $1\cdot 10^{-4}$  &0.000849357 & 0.0008491& $2\cdot 10^{-7}$ \\
        \hline
        $D$ & $\mathcal{S}_{dir,2}$& $\mathcal{S}_{LHF,2}$ & 
        $\Delta \mathcal{S}_{dir,2}$& $\mathcal{H}_{dir,2}$ &$\mathcal{H}_{LHF,2}$&$\Delta \mathcal{H}_{dir,2}$\\
        \hline
        10 & 0.884620 & 0.884645 & $2\cdot 10^{-5}$ & 0.457704 &0.457705 & $1\cdot 10^{-6}$  \\
        1   & 0.6303281 &0.6303235  &$5 \cdot 10^{-6}$ & 0.2109513& 0.2109511 & $1\cdot 10^{-7}$ \\
        0.1  &0.08058915 &0.08058649 &$3\cdot 10^{-6}$ &0.0008230 &0.0008230367 & $3\cdot 10^{-8}$ \\
        \hline
    \end{tabular}
    \caption{Comparison of various computed values of the steepness and energies of the first two non-trivial extremizers of the energy in depths 10, 1, and 0.1. The first set of three columns corresponds to the directly computed steepness of the first or second extremizer of the energy, the approximation of this steepness by the asymptotics of Longuett-Higgins and Fox, and the error between these values.  The last three columns show the value of the extremizer itself in the same way.}
    \label{tab:ExtremaComparisons}
\end{table}

\begin{align}
   R_n = \frac{\mathcal{S}_{\mathcal{H},n}-\mathcal{S}_{c,n-1}}{\mathcal{S}_{c,n}-\mathcal{S}_{c,n-1}}. \label{eq:Rndef}
\end{align}

\no The ratio $R_n$ measures how far between two extremizers of the speed the $n$th extremizer of the energy is located. To approximate this limit we note that Longuett-Higgins and Fox's perturbation parameter $\epsilon$ tends to zero as the steepest wave is approached. Thus, for large $n$, we approximate $R_n$ by
\begin{align}
    R_n\sim \frac{\mathcal{S}_{\lim}+d_1\epsilon_{H,n}^2-(\mathcal{S}_{\lim}+d_1\epsilon_{c,n-1}^2)}{\mathcal{S}_{\lim}+d_1\epsilon_{c,n}^2-(\mathcal{S}_{\lim}+d_1\epsilon_{c,n-1}^2)}=\frac{\epsilon_{H,n}^2-\epsilon_{c,n-1}^2}{\epsilon_{c,n}^2-\epsilon_{c,n-1}^2}.\label{eq:R_nApprox1}
\end{align}
Using (\ref{eq:epsSpeed}) and (\ref{eq:epsEnergy}), 
\begin{align}
    \epsilon_{c,n} = \epsilon_{c,n-1}e^{-\frac{\pi}{3\mu}},\quad\text{and}\quad \epsilon_{\mathcal{H},n} = \epsilon_{\mathcal{H},n-1}e^{-\frac{\pi}{3\mu}}. \label{eq:epsilonRecursion}
\end{align}
To justify (\ref{eq:epsilonRecursion}), note that as $n\to \infty,$ $k \to -\infty$ in (\ref{eq:epsSpeed}) and (\ref{eq:epsEnergy}) and $\epsilon \to 0$ so the steepest wave is approached. We see that 
\begin{align}
    \lim_{n\to\infty} R_n &\sim \lim_{n\to\infty} \frac{\epsilon_{\mathcal{H},n}^2-\epsilon_{c,n-1}^2}{\epsilon_{c,n}^2-\epsilon_{c,n-1}^2}\nonumber \\
    &=\frac{e^{-\frac{2\pi}{3\mu}(n-2)}\left(\epsilon_{\mathcal{H},2}^2-\epsilon_{c,1}^2\right)}{e^{-\frac{2\pi}{3\mu}(n-2)}\left(\epsilon_{c,2}^2-\epsilon_{c,1}^2\right)}\nonumber\\
    &=\frac{e^{-\frac{2\pi}{3\mu}}\epsilon_{\mathcal{H},1}^2-\epsilon_{c,1}^2}{\epsilon_{c,1}^2\left(e^{-\frac{2\pi}{3\mu}(n-2)}-1\right)}\nonumber\\
    &=\frac{e^{-\frac{2\pi}{3\mu}}\frac{\epsilon_{\mathcal{H},1}^2}{\epsilon_{c,1}^2}-1}{e^{-\frac{2\pi}{3\mu}}-1}.\label{eq:RnApprox}
\end{align}

Thus, we compute the limiting ratio using just the ratio of the perturbation parameters corresponding to the first extremizers of the energy and speed. We could have used steeper extremizers as well, but the first one is most accessible. %\textcolor{red}{In future work, we refine} this by using steeper extremizers. 
We compute this limiting ratio in different depths and {present results} in Figure~\ref{fig:LimitingRatioFit}. Despite the noise introduced by the Longuett-Higgins and Fox asymptotics in a regime far beyond their validity, we can fit a rescaled, shifted power of the dispersion relation $\omega^2 = gk\tanh(kd)$ to this data, with an exponent whose confidence interval includes 1.

\begin{figure}
    \centering
    \includegraphics[width=\linewidth]{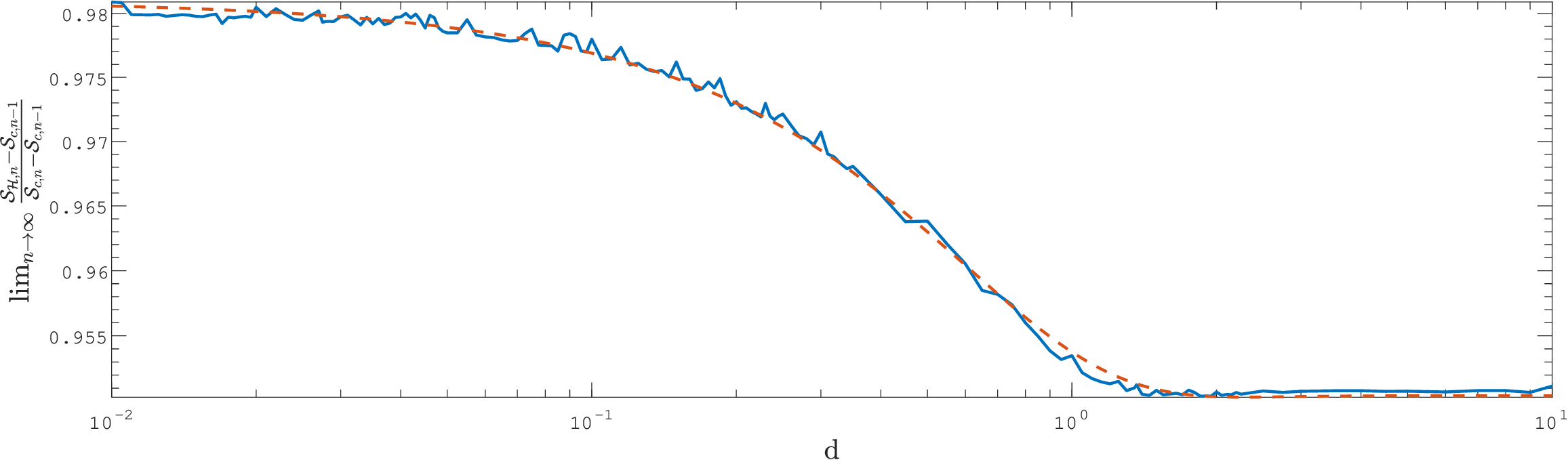}
    \vspace*{0.1in}
    \caption{Fit of the value of the approximation to the limit $\lim_{n\to\infty}R_n$, see \eqref{eq:Rndef}, as a function of depth to a rescaled shift of an inverse power of the modified dispersion relation $\tilde{\omega}(d)$. The relative error of the fit is on the order of $10^{-3}$ when we set the inverse power to $-1$. Computing these ratios with more waves may reduce the noise in our data and allow for more accuracy and confidence in this fit. The solid blue line is the computed value of the limiting ratios, whereas the dashed red line is a fit $0.98-0.3\tilde{\omega}(0.9d)^{-1}.$}
    \label{fig:LimitingRatioFit}
\end{figure}

\section{Branch Point Singularities} \label{sec:Pole Structure}

We study the singularity structure of the analytic continuation of the Stokes waves and compare it to that of infinite depth waves \cite{crew2016new,dyachenko2014complex,dyachenko2016branch,grant1973singularity,longuet1977theory,lushnikov2016structure,schwartz1974computer}. To our knowledge most of these works have not been extended to finite depth. In infinite depth, knowledge of the singularities has not only been used to better understand the formation of a $120^\circ$ corner but also to develop more efficient methods to compute and study these waves. Similarly\textit{,} it is of great interest to understand the singularity structure in finite depth.  In particular, our goal is to generalize the results of Dyachenko \textit{et al} \cite{dyachenko2014complex} to the finite-depth case. 

Using conformal variables, Dyachenko \textit{et al} \cite{dyachenko2014complex} deform the contour of integration in the integral definition of the Fourier coefficient to encircle the singularity above the fluid. The decay of these coefficients directly relates to the type of singularity above the fluid. In particular, if the singularity has a dominant contribution

\beq
z \sim c_1(w-iv_c)^{\beta},
\eeq 

\no then the Fourier spectrum of $z$ decays as

\begin{align}
|z_k| \sim |k|^{-1-\beta}\exp(-|k|v_c),\quad k \to -\infty.\label{eq:branchAbove}
\end{align}

\no Dyachenko \textit{et al} \cite{dyachenko2014complex} found a square root-type branch point above the fluid, and determined that $v_c$ decayed as
\begin{align}
    v_c\sim \left( \mathcal{S}_{\lim}-\mathcal{S} \right)^{\delta}, \label{eq:vcDecay}
\end{align}
where $\delta = 1.48 \pm 0.03$ includes $3/2$ within its confidence interval. 

A natural change from infinite depth comes from the presence of singularities beneath the conformal domain. Recall that the velocity potential $\phi$ satisfies the Laplace Equation (\ref{eq:Laplace}) within the bulk of the fluid with a Neumann boundary condition (\ref{eq:noPenetration}) on the bottom boundary. Thus, the analytic continuation of solutions beneath this boundary satisfies a mirror principle. In conformal variables, $\phi(u,v)=\phi(u,-2h-v)$. Thus any singularity above the conformal domain is mirrored beneath the conformal domain as shown in Figure~\ref{fig:wavesWithSingularities}.

\begin{figure}
    \centering
    \includegraphics[width=\linewidth]{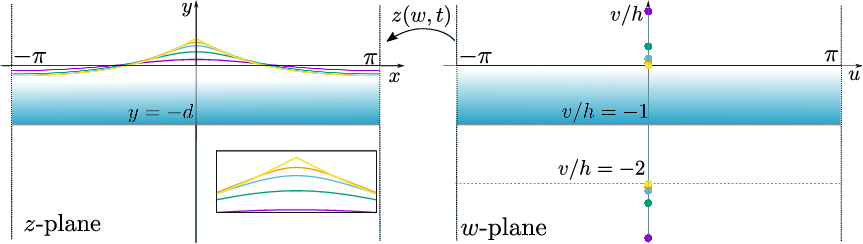}
    \caption{Waves of increasing steepness in depth 1, and the conformal mapping. Colored dots correspond to colored waves, showing the approach of singularities above and below the fluid to the conformal domain. The vertical axis for the right figure displays $x/h$, to account for different conformal depths for steeper waves in constant physical depth.}
    \label{fig:wavesWithSingularities}
\end{figure}

To leading order, a singularity beneath the fluid has the form
    
\begin{align}
z = c_2(w+iv_c')^{\beta'},\quad v_c'>2h.
\end{align}

\no If we deform the contour of the integral
    \begin{align}z_k = \frac{1}{2\pi}\int_{-\pi}^\pi z(u)e^{iku}du,\quad k>0,\end{align}
    around a singularity below the fluid, we find the asymptotic form
    \begin{align}
        |z_k| \sim |k|^{-1-\beta}\exp(-|k|v_c'),\quad k \to \infty.\label{eq:branchBeneath}
    \end{align}
    To determine $v_c'$ and $\beta'$ for the singularity beneath the fluid, we use Stokes waves computed with 64 bits of precision and fit the Fourier spectrum to (\ref{eq:branchAbove}-\ref{eq:branchBeneath}) using Julia's non-linear fitting routine, with accuracy $10^{-32}$. For every wave, it is confirmed that the locations of the singularities above and below the fluid are directly linked through the conformal depth $h$, with $v_c'+v_c+2h$ on the order of $10^{-3}$ or less for every wave checked. A sample fit of the decay of the spectrum of a relatively steep wave in depth 0.2 is shown in Figure~\ref{fig:BranchFit}. Since $\beta = \beta'=1/2$, these fits show that there are square-root type branch points both above and below the conformal domain. This verifies the predictions of Grant \cite{grant1973singularity} and Tanveer \cite{tanveer1991singularities} about the singularity above the fluid, as well as the use of the mirror principle to understand the singularity below the fluid for finite-depth Stokes waves.

    \begin{figure}[h]
    \centering
    \includegraphics[width=\linewidth]{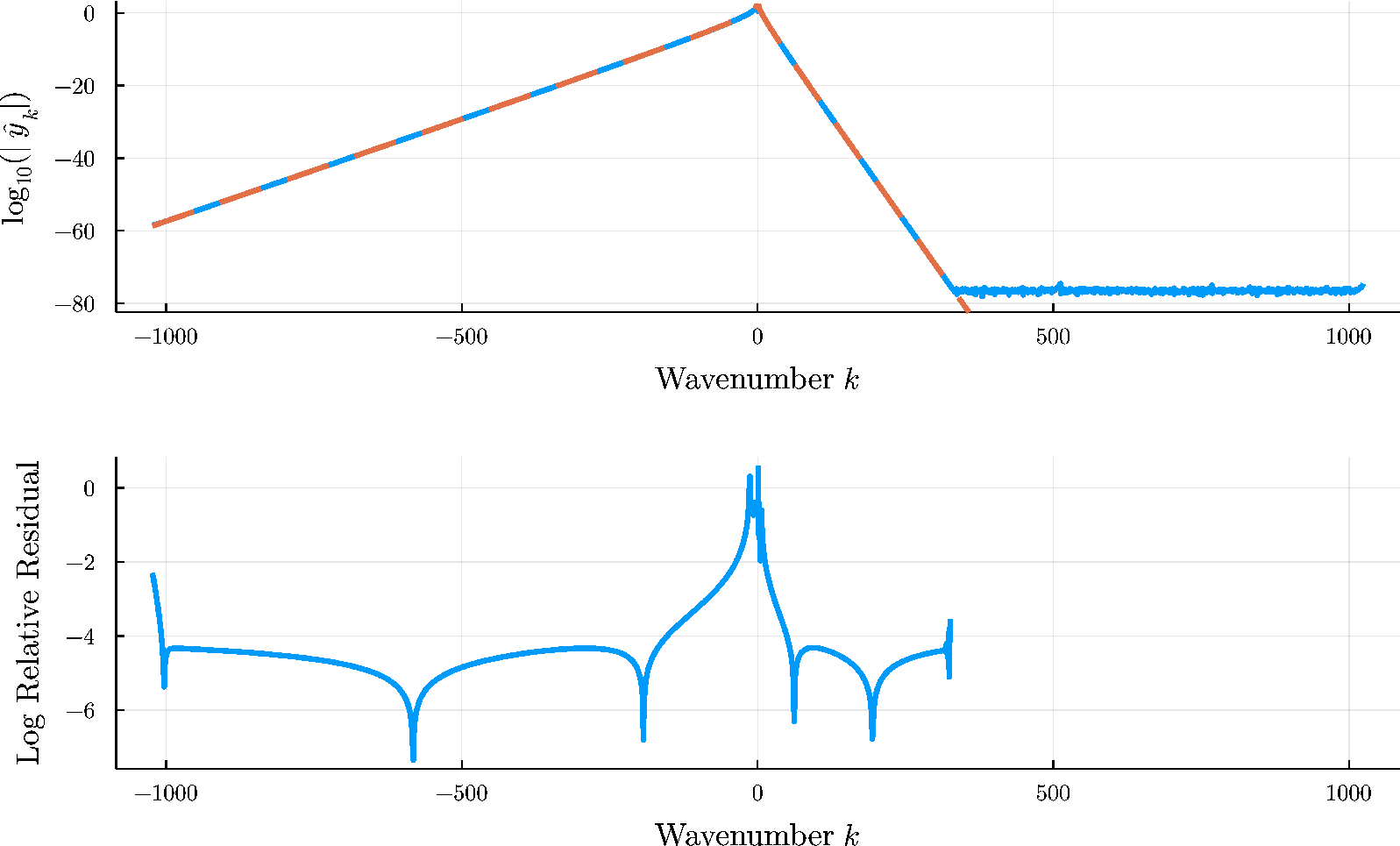}
    \caption{Top: Fit of the asymptotics given in  (\ref{eq:branchAbove}) and (\ref{eq:branchBeneath}). Bottom: relative residuals on a semi-log scale. Note the high degree of accuracy for $k$ sufficiently far from zero. A portion of the relative error between the positive Fourier coefficients and the fit is cut off, as the coefficients are at the level of noise, leading to growing relative errors in this regime despite a good fit. The solid blue line shows the Fourier modes of the surface elevation $y$ and the dashed red line shows the fits.}
    \label{fig:BranchFit}
\end{figure}
    
    We repeat the calculations of {Dyachenko \textit{et al}} \cite{dyachenko2014complex} regarding the branch point above the fluid domain of various extreme waves in double precision. Figure~\ref{fig:BranchDecay} demonstrates that the singularity above the fluid in depth 0.01 approaches the conformal domain in the same manner as described by Dyachenko \textit{et al} \cite{dyachenko2014complex,dyachenko2016branch} and Lushnikov \cite{lushnikov2016structure}. In Figure~\ref{fig:vcEvolution} we show that this rate of approach is effectively constant as a function of depth. The similarity of these results is encouraging for an exploration of the structure of the Riemann sheet structure connected to this branch point \cite{crew2016new,lushnikov2016structure}, or the development of more efficient methods for the computation of waves in a fixed finite depth \cite{Dyachenko2023Almost,lushnikov2017new,tanveer1991singularities}. Further investigations of the behavior of these branch points in the shallow water regime is left for future work.

\begin{figure}[h]
    \centering
    \includegraphics[width=\linewidth]{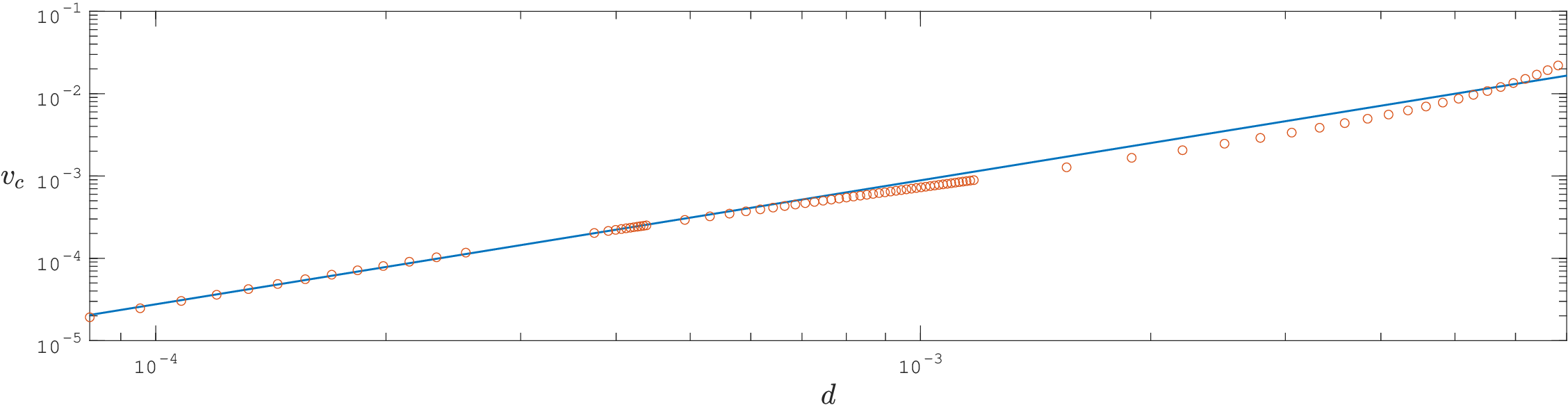}
    \caption{Fit of singularity location above the fluid $v_c$ as a power of the difference in steepnesses $\mathcal{S}_{\lim}-\mathcal{S}$ in depth 0.01. In this shallow water as in infinite depth, a rate of decay of approximately 1.5 is predicted. The relative residual of the error between our fit and the data is bounded by $10^{-1}$ for extreme waves suggesting this is a first-order term. The red circles are the computed branch point locations, and the blue line is a fit.}
    \label{fig:BranchDecay}
\end{figure}

\begin{figure}[h]
    \centering
    \includegraphics[width=\linewidth]{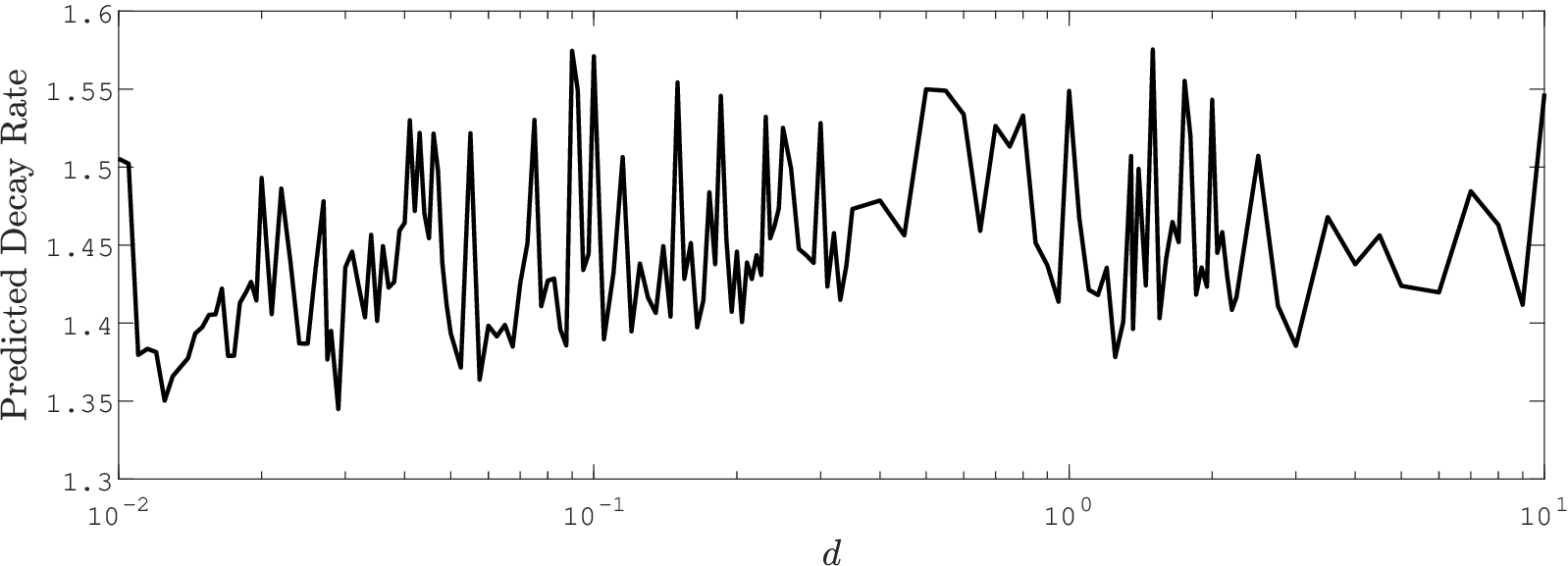}
    \caption{Change in the exponent $\delta$ of (\ref{eq:vcDecay}), the location of the singularity above the fluid as depth varies. Note that all predicted exponents are close to 1.5, as in infinite depth. Noise is largely due to which depths have more or fewer computed waves.}
    \label{fig:vcEvolution}
\end{figure}

\section*{Acknowledgements}
The authors thank Sergey Dyachenko for helpful conversations. The work of EB was generously funded in part by the Rodney Mason Wan Fellowship and the ARCS Foundation. AS aknowledges the Pacififc Institute of Mathematical Sciences and Simons Foundation for their support.

\section*{Declaration of Interests} The authors report no conflicts of interest. 

\bibliographystyle{plain}
%\nocite{*}
\bibliography{presentation}

\end{document}